\documentclass[aps,prl,reprint,a4paper,superscriptaddress,longbibliography,amsmath,amssymb,floatfix]{revtex4-2}

\usepackage{hyperref}
\hypersetup{
    colorlinks=true,
    citecolor=blue,
    linkcolor=blue,
    urlcolor=blue,
    pdftitle={Normalizing flows for all-order QED corrections in lattice field theory},
    pdfauthor={Nils Hermansson-Truedsson, Gurtej Kanwar},
    pdfpagemode=FullScreen
}
\usepackage[explicit]{titlesec}
\titlespacing{\paragraph}{1em}{0pt}{2pt}[]
\usepackage{xcolor}
\usepackage{graphicx}
\graphicspath{{./Figures/}}

\newcommand{\qedl}{\text{QED$_\text{L}$}}

\newcommand{\sfull}{S(\phi , A; e)}
\newcommand{\sint}{S_{\mathrm{int}}(\phi, A; e)}
\newcommand{\sintdern}[1]{S_{\mathrm{int}}^{( #1 )}(\phi, A)}
\newcommand{\sintdernp}[1]{S_{\mathrm{int}}^{( #1 )}(\phi', A')}
\newcommand{\sprime}{\delta _{e^2} e^{-\sint}}
\newcommand{\sbar}{\bar{S}(\phi , A ; e)}

\newcommand{\obsiso}{\langle \mathcal{O} \rangle _0}
\newcommand{\obspe}{\langle \mathcal{O} \rangle _{e}}

\newcommand{\obsprime}{\langle \mathcal{O} \sprime \rangle _{0,c} }

\newcommand\bsquare{\blacksquare}

\begin{document}

\title{Normalizing flows for all-orders QED corrections in lattice field theory}

\author{Nils Hermansson-Truedsson}
\email[]{nils.hermansson-truedsson@ed.ac.uk}
\author{Gurtej Kanwar}
\email[]{gkanwar@ed.ac.uk}
\affiliation{Higgs Centre for Theoretical Physics, School of Physics and Astronomy, The University of Edinburgh, James Clerk Maxwell Building,
Peter Guthrie Tait Road,
Edinburgh,
EH9 3FD}

\date{\today}

\begin{abstract}
This work develops a framework to apply normalizing-flow transformations of field configurations for all-orders Quantum Electrodynamics (QED) corrections in lattice field theory. 
This opens a new possibility to determine all-order corrections without the need for additional Monte Carlo sampling, generally bypassing the complexity in Wick-contraction diagrams needed at fixed order.  The new method is applied to lattice scalar QED in two, three, and four spacetime dimensions, using both analytical and machine-learned flows, with both approaches yielding estimates with significantly reduced variance with respect to standard methods.
It is further shown that flows can be trained using small lattice geometries and subsequently evaluated on much larger lattice geometries while maintaining good efficiency. A generalization to theories with fermions is envisaged, suggesting a path to applications in challenging field theories including lattice Quantum Chromodynamics.
\end{abstract}

\maketitle

\paragraph{Introduction}---\,Electromagnetic corrections and strong isospin-breaking effects from light-quark mass differences are needed for a growing number of low-energy precision tests of the Standard Model~\cite{FlavourLatticeAveragingGroupFLAG:2024oxs}. To date, these effects have been evaluated for a variety of observables from first-principles lattice quantum chromodynamics (QCD) with explicit quantum electrodynamics (QED) corrections~\cite{BMW:2014pzb,deDivitiis:2013xla,Clark:2022wjy,Frezzotti:2022dwn,RCstar:2022yjz,Altherr:2026pux,Carrasco:2015xwa,Giusti:2017dwk,DiCarlo:2019thl,Boyle:2022lsi,Christ:2023lcc,Giusti:2017jof,RBC:2018dos,Giusti:2017dmp,Giusti:2019xct,Aoyama:2020ynm,Borsanyi:2020mff,Biloshytskyi:2022ets,Chao:2023lxw,Boccaletti:2024guq,Djukanovic:2024cmq,Parrino:2025afq,RC:2025zoa,Aliberti:2025beg,Bruno:2026kqf}.
In this work we develop a new technique to determine electromagnetic corrections in lattice field theory with reduced variance, which has the potential to enable determining currently unknown quantities in QCD+QED of phenomenological interest, such as the neutron axial charge~\cite{Tomalak:2026wks,Cirigliano:2024nfi,Gorchtein:2023srs}.

We propose to construct lattice normalizing flows~\cite{tabak2010,tabak2013,rezende2015variational,Albergo:2019eim} to deterministically transform field configurations from an originally uncoupled theory $(e = 0)$ to the fully interacting theory, including QED dynamics at all orders.
Similar techniques have been recently applied to reduce variance of inserted operators or correlation functions in lattice QCD~\cite{Abbott:2024kfc,Abbott:2026ylv}. Here, this method is extended and specialized for the context of QED corrections. A key innovation is the application of flows in way that exactly cancels the $O(e)$ contributions to the variance of measured observables. This is explored first by constructing an analytical normalizing flow with this property, then training and applying machine-learned flows in a setup that enforces this cancellation and further improves variance over the analytical flow.
Notably, this approach does not require new Monte Carlo sampling to be performed, making it directly applicable to existing ensembles which are often generated at great cost. This stands in contrast to several comparable methods, such as all-orders QCD+QED using $C^*$ boundary conditions~\cite{Lucini:2015hfa,RCstar:2022yjz,RC:2025zoa} or stochastic automatic differentiation for isospin-breaking beyond leading order~\cite{Albandea:2026cqh}.
This also allows QED effects to be precisely isolated by jointly analyzing correlated samples from the uncoupled and fully interacting theory. Finally, the flow approach evaluated at all orders has the computational benefit of avoiding demanding Wick contractions~\cite{BMW:2014pzb,Endres:2015gda,Boyle:2017gzv,RC:2025zoa,Altherr:2026pnt}.

To test the method, a numerical evaluation is presented for observables in scalar QED in two, three, and four dimensions.
Our key numerical results are operator expectation values calculated in scalar QED to all orders in the electromagnetic coupling. 
Representative fits to extract the electromagnetic mass shift are also compared to results from traditional lattice methods~\cite{Davoudi:2018qpl,Bijnens:2019ejw}, and in all cases the flow method provides controlled estimates at fixed statistics for much larger volumes than is accessible by direct evaluation. 
As there is no unique way to regularize QED in a finite volume~\cite{Kronfeld:1990qu,Wiese:1991ku,Kronfeld:1992ae,Polley:1993bn,Duncan:1996be,Duncan:1996xy,Hayakawa:2008an,Asmussen:2016lse,Endres:2015gda,Feng:2018qpx,Davoudi:2018qpl,Clark:2022wjy,Biloshytskyi:2022ets,DiCarlo:2025uyj}, all numerical demonstrations here are performed for a typical choice of scheme, $\qedl$~\cite{Hayakawa:2008an,Davoudi:2018qpl}. However, the normalizing flow approach is generally applicable also for other finite-volume QED regularizations, and transferring flow models to these other choices represents interesting future work.
These results open the door to analogous gains via machine-learned normalizing flows for lattice QCD+QED.

\paragraph{Flows for all-orders QED effects}---\, Starting from a theory with lattice action $S_0(\phi)$ and field content generically written as $\phi$, QED effects can be included by introducing the photon field $A_\mu(x)$, the free photon action $S_A(A)$, and the QED interaction term $S_{\rm int}(\phi, A; e)$ with coupling $e$. The free photon action typically introduces a distribution that is easy to sample, e.g., a Gaussian action which can be sampled exactly. As such, we will take as our starting point field configurations sampled according to the uncoupled theory described by the distribution
\begin{equation} \label{eq:prior-density}
    r(\phi, A) = \frac{1}{Z_0} e^{-S_0(\phi) -S_A(A)}
\end{equation}
normalized by $Z_0 \equiv \int d[\phi] \, d[A] \, e^{-S_0(\phi) -S_A(A)}$.
The factorized distribution allows configurations of $A$ to be generated independently of $\phi$, for example being produced on-demand after an ensemble of fields $\phi$ has already been sampled.

Here we propose to use a normalizing flow to transform the uncorrelated prior distribution \eqref{eq:prior-density} to the all-orders interacting theory, described by the distribution
\begin{equation} \label{eq:target-density}
    p(\phi, A) = \frac{1}{Z_e} e^{-S_0(\phi) - S_A(A) - S_{\rm int}(\phi, A; e)},
\end{equation}
where $Z_e \equiv \int d[\phi] \, d[A] \, e^{-S_0(\phi) - S_A(A) - S_{\rm int}(\phi, A; e)}$.
A normalizing flow transformation $f$ is any a diffeomorphism with tractable Jacobian determinant~\cite{rezende2015variational}, which here maps samples drawn from $r(\phi, A)$ to new field configurations,
\begin{equation}
    f: (\phi, A) \mapsto (\phi', A').
\end{equation}
The transformed fields $(\phi', A')$ follow a new distribution described by the change of measure formula
\begin{equation} \label{eq:flow-density}
    q(\phi', A') = r(\phi, A) |\det J_f|^{-1},
\end{equation}
where $J_f$ is the Jacobian of $f$.
By optimizing the flow $f$ such that $q(\phi', A') \approx p(\phi', A')$, this transformation provides a way to approximately map from an uncoupled ensemble to fields distributed according to the all-orders interacting theory. Even for an imperfect flow, exact reweighting factors can be computed as
\begin{equation}  \label{eq:what}
    \hat{w}(\phi', A') \equiv \frac{e^{-S_0(\phi') - S_A(A') - S_{\rm int}(\phi', A'; e)}}{e^{-S_0(\phi) - S_A(A)}} |\det J_f|.
\end{equation}
These unnormalized reweighting factors provide an unbiased estimator of the ratio of partition functions, $\left< \hat{w} \right>_0 = Z_e / Z_0$, leading to the normalized weights
\begin{equation} \label{eq:w}
    w(\phi', A') \equiv \hat{w}(\phi', A') / \left< \hat{w} \right>_0.
\end{equation}
Here expectation values $\left< \cdot \right>_0$ are taken with respect to $(\phi, A)$ drawn from the \emph{non-interacting} theory, with the transformed fields $(\phi', A')$ computed deterministically from these samples by the application of $f$.
Finally, observables in the all-order interacting theory can likewise be evaluated as
\begin{equation} \label{eq:flowed-obs}
    \left< \mathcal{O} \right>_e = \frac{\left< \mathcal{O}(\phi', A') \, \hat{w} \right>_0}{\left< \hat{w} \right>_0} = \left< \mathcal{O}(\phi', A') w(\phi', A') \right>_0.
\end{equation}
A given flow $f$ only needs to be optimized and evaluated once and can be used to evaluate many choices of observables $\mathcal{O}$.

Recent works have considered a number of machine-learning parameterizations of $f$ which are then optimized to reproduce the desired change of measure in lattice field theory contexts; for recent reviews, see Refs.~\cite{Cranmer:2023xbe,Kanwar:2024ujc,Wenger:2026sjp}. This work applies both an analytical parameterization, detailed below, and a machine learning architecture. As we work with a gauge-fixed $\qedl$ prescription, the gauge symmetry does not need to be explicitly treated, and here the machine learning model is parameterized using a residual U-net architecture~\cite{chen2019residual,ronneberger2015u}. In all cases, the machine-learned flows are optimized using the Kullback-Leibler divergence~\cite{Kullback:1951zyt} with path gradients~\cite{vaitl2022gradients} for a fixed number of optimization steps. The architecture and training method are detailed in the supplement.

For charge-conjugation-even observables, leading corrections begin at $O(e^2)$, while a formal expansion of the flow transformation must begin at $O(e)$ to fully reproduce \eqref{eq:target-density}. This implies that \eqref{eq:flowed-obs} includes fluctuations at $O(e)$ that dominate the leading $O(e^2)$ effects. A standard approach for such observables is to average their value over correlated evaluations at couplings $\pm e$~\cite{BMW:2014pzb}. Typically this is only possible in the electroquenched case, because it is impossible to flip the sign of $e$ after sampling the dynamical theory. However, with the flow method this is possible: we exploit charge-conjugation symmetry to translate a flow $f$ for coupling $e$ to a related flow $f_-$ for coupling $-e$ according to
\begin{equation}
\begin{aligned}
    f_-(\phi, A) &\equiv (\phi'_-, A'_-) = (\bar{\phi'}, -\bar{A'}), \\
    (\bar{\phi'}, \bar{A'}) &= f(\phi, -A).
\end{aligned}
\end{equation}
The associated reweighting factors $w_-(\phi'_-, A'_-)$ are computed as in \eqref{eq:what} and \eqref{eq:w} with $e \to -e$.
From this, we define the improved observable
\begin{equation}
\begin{aligned}
    \obspe &= \frac{1}{2} \big\langle \mathcal{O}(\phi', A') w(\phi', A') \\
    &\qquad +\mathcal{O}(\phi'_-, A'_-) w_-(\phi'_-, A'_-) \big\rangle.
\end{aligned}
\end{equation}

\paragraph{Scalar QED}---\,To demonstrate our flowed approach we consider lattice scalar QED on an $N_d$-dimensional lattice $\Lambda _{N_d}$ with geometry $\hat{L}^{N_d - 1} \times \hat{T}$. To regulate the zero-momentum modes of photons we employ the $\qedl$ prescription~\cite{Hayakawa:2008an,Davoudi:2018qpl}.
Scalar QED contains a charged scalar field $\phi$ and the $U(1)$ gauge field $A_\mu$, with dynamics described in Feynman gauge by the lattice action~\cite{Davoudi:2018qpl,Bijnens:2019ejw}
\begin{align}\label{eq:lagrangian}
S(\phi, A; e) &= S_0(\phi) + S_A(A) + \sint
     \, ,
\end{align}
where
\begin{align}
    S_A(A) &= - \frac{1}{2}\, A_{\mu}(x) \delta ^2 A_{\mu} (x) \, , \label{eq:action-A} \\
    S_0(\phi) + \sint &= \sum_{x \in \Lambda_{N_d}} \phi^\dagger(x) \Delta \phi(x) \, . \label{eq:action-phiA}
\end{align}
The discretized derivative operators above are given by
\begin{align}
    \Delta & = m^2_0 + \frac{1}{a^2}\sum _{\mu }\left( 2-e^{ieaA_\mu} \tau _\mu -\tau_{-\mu}e^{-ieaA_{\mu}}\right)  \, ,
    \\
    \delta ^2 & = \frac{1}{a^2}\, \sum _{\mu} (\tau_\mu +\tau_{-\mu}-2)
    \, ,
\end{align}
where the translation operator $\tau _\mu$ transports a field $\chi$ according to $\tau _\mu \, \chi(x) = \chi(x+ a\hat{\mu})$.
The operator $\Delta $ can be expanded in powers of $e$ as $\Delta = \sum _{n=0}^{\infty} e^{n}\, \Delta _{n}$,
which naturally allows $S_0(\phi) = \sum_x \phi^\dagger \Delta_0 \phi$ to be isolated in \eqref{eq:action-phiA}, with $\sint = \sum _{n=1}^{\infty} e^n  \, \sintdern{n}  $
given by the remaining terms.

For the $\qedl$ regularization, the path integral includes a projection that removes spatial zero-momentum modes of the photon on every time slice $t$ according to
$\prod_t \delta(\sum_{\mathbf{x}} A_\mu(t, \mathbf{x}))$,
where  $\mathbf{x}\in \Lambda ^{N_d -1}$ is the spatial coordinate. For the flow approach, this constraint is included in the definition of the prior distribution \eqref{eq:prior-density} and must be respected by the flow transformation $f$; in practice this is implemented by applying a linear projector $P_{\qedl}[\cdot]$, which maps into the $\qedl$ subspace.

The specific form of the scalar QED interaction allows the definition of an \emph{analytical} normalizing flow for this theory. Because the $O(e)$ interaction can be written as
\begin{align}
    \sintdern{1} &= \sum_{x \in \Lambda_{N_d}} \sum_{\mu} A_\mu(x) \, j_\mu(x), \\
    j_\mu(x) &\equiv \frac{2}{a} \, \mathrm{Im}(\phi^\dagger(x) \tau_\mu \phi(x)),
\end{align}
the total action at this order is a simple Gaussian in $A$.  This motivates constructing the following linear normalizing flow acting only on $A$,
\begin{equation} \label{eq:linear-flow}
{A'}^{\mathrm{Lin.}}_\mu(x) \equiv A_\mu(x) + e P_{\qedl}\left[(\delta^{2})^{-1} j_\mu(x) \right],
\end{equation}
where $(\delta^2)^{-1}$ is the pseudo-inverse of the discretized Laplacian operator (i.e., dropping the global zero-mode), and the Jacobian $J_f$ of this transformation is triangular with $\det J_f = 1$. This transformation induces a change of measure that matches the target distribution exactly at leading order.

\paragraph{Numerical comparisons}---\,We compare the expectation value and variance of $\obspe$ in the interacting theory using several methods.
First, we employ the linear flow in \eqref{eq:linear-flow}. While it is not straightforward to analytically generalize this flow to theories with fermions, this transformation provides initial evidence that normalizing flows with significant benefits can be found.
Second, we use machine-learned normalizing flows
with models trained in a small, cubically symmetric volume with geometry either $\hat{L} = \hat{T} = 8$ ($N_d = 2,3$) or $\hat{L} = \hat{T} = 4$ ($N_d = 4$).
For both the analytical and machine-learned flows, these models are evaluated for a range of larger volumes with rectangular geometry $\hat{T} = 2 \hat{L}$
for
the full all-orders action $\sfull$ with $m_0 = 1$ and $e=0.3$. Finally, as a baseline for comparison, we also perform brute-force stochastic computations of the non-perturbative matrix elements appearing. For the all-orders case, this corresponds to reweighting expectation values $\obspe = \left< \mathcal{O} e^{-\sint} \right>_0 / \left< e^{-\sint} \right>_0$.

As a representative example, we focus on operators $\mathcal{O} = \phi (t,\boldsymbol{0}) \, \phi ^\dagger (0,\boldsymbol{0})$ from which a zero-momentum, two-point correlation function is defined as
\begin{align}\label{eq:corre}
   C_{e}(t) =  \frac{1}{\hat{L}^{N_d -1}}\sum _{\mathbf{x}} \langle \phi (t,\boldsymbol{0}) \, \phi ^\dagger (0,\boldsymbol{0}) \rangle _e \, . 
\end{align}
This correlation function allows extracting the mass of the scalar field, $m_{e}$, by analyzing its temporal dependence.
In particular, denoting the free theory correlator by $C_{0}(t)$,
the mass shift can be extracted by studying the spectral representation of the ratio
\begin{equation}
    C_e(t)/C_0(t) \approx \frac{A_e}{A_0} e^{-\Delta m_e t} + \dots  \, ,
\end{equation}
where the ellipsis indicates excited state and finite-$T$ effects
which are suppressed for $0 \ll t \ll T/2$.
For visualization, we also define the effective mass shift $\Delta m_{\rm eff} = -\partial _{t} \log C_e(t)/C_0(t)$ which converges to $\Delta m_e$ in the same region.
In $N_d=2,3$ the mass shift is infrared divergent with the lattice volume, but for the numerical comparisons at the fixed volumes used here, we do not need to address this issue. 
As discussed above, order-$e$ noise is mitigated by averaging  $C_{e}(t)$ over coupling values $\pm e$ in all methods.

\begin{figure}
    \centering
    \includegraphics{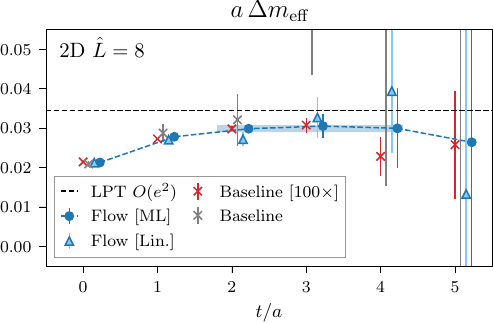}
    \caption{The effective mass as a function of time, obtained from the derivative of the logarithm of the correlator ratio, for $N_d = 2$ and $\hat{L} = 8$. }
    \label{fig:effective_mass2d}
\end{figure}

\begin{figure}
    \centering
    \includegraphics{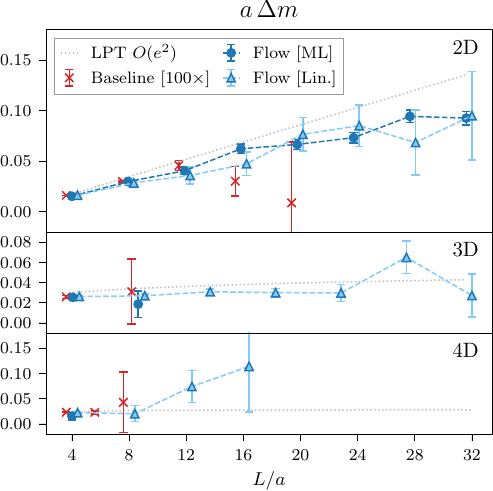}
    \caption{The fitted all-orders mass shift $\Delta m_e$ as a function of the lattice size, $\hat{L} \equiv L/a$, for $N_d = 2, 3, 4$ with geometry $\hat{T} = 2\hat{L}$. The machine-learned and analytical flows are compared against a baseline with $100\times$ more statistics and the LPT $O(e^2)$ result.
Only data with uncertainties $\lesssim 100\%$ are plotted.}
    \label{fig:delta_m}
\end{figure}

As a first step in the comparison, the correlator ratio $C_e/C_0$ and the corresponding effective mass $\Delta m_{\rm eff}$ are compared for the range of lattice geometries studied using ensembles of size $10^5$. A representative example is shown in Fig.~\ref{fig:effective_mass2d} for $N_d = 2$ at a fixed lattice size $\hat{L} = 8$, $\hat{T} = 2 \hat{L}$. A clear hierarchy in precision can be observed across the methods at fixed statistics: while all methods suffer from the signal-to-noise problem in the underlying correlator, the ML flow provides the most precise estimates, the analytical linear flow provides comparable estimates with uncertainties approximately a factor of two larger, and the baseline stochastic estimates have much larger uncertainties. A baseline evaluation with $100$ times more statistics is also shown for reference, verifying the correctness of the flow results. For reference, the order-$e^2$ finite-volume lattice perturbation theory (LPT) mass shift is also shown (details provided in the supplement), demonstrating that the all-orders flow data is able to resolve the subleading shift consistent with $O(e^4)$ and higher effects.
To compare the effects on final extracted quantities, the correlator ratios $C_e / C_0$ are then fitted to the single-exponential form $A e^{-\Delta m \, t}$ within a fixed range in $t/a$ for each value of $N_d$ and $\hat{L}$. While this fitting procedure certainly is still affected by excited-state effects, the consistent fit windows allow a direct comparison in precision of derived quantities between the methods, as summarized across all values of $\hat{L}$ and $N_d = 2, 3, 4$ in Fig.~\ref{fig:delta_m}.
Notably, a single trained ML model is directly used to evaluate at all choices of $\hat{L}$ for each $N_d$, with evaluation geometries being up to four times larger in each extent than the training geometry. Because of significant noise in the stochastic baseline, the baseline with higher statistics is shown for comparison. For clarity of visualization, all results are truncated when relative uncertainties grow larger than $\simeq 100\%$. 
Here LPT results are also plotted for reference, and again a notable subleading correction to the $O(e^2)$ result is possible to resolve with the flow results. We note however that quantitatively extracting this subleading shift would require further statistics and control of excited-state effects.

\begin{figure}
    \centering
    \includegraphics{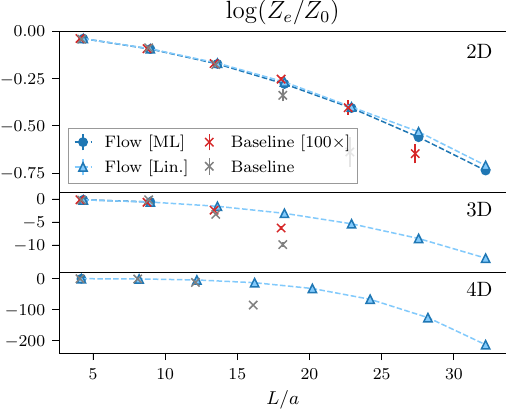}
    \caption{Comparison of $\log(Z_e/Z_0)$ as a function of lattice size $\hat{L} \equiv L/a$ for $N_d = 2, 3, 4$. All results correspond to $\hat{T} = 2\hat{L}$.}
    \label{fig:logz}
\end{figure}

\begin{figure}
    \centering
    \includegraphics{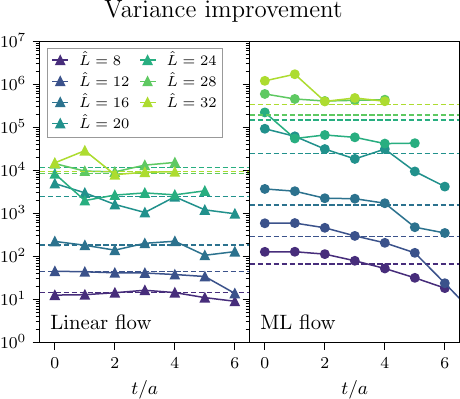}
    \caption{Factor by which the variance of $C_e/C_0$ is reduced when estimated using the linear flow (triangles, left) or the machine-learned flow (circles, right) versus the standard stochastic estimate for $N_d = 2$. The factor improvement in the variance of $\log{Z_e/Z_0}$ is also shown in dashed lines. Results are plotted for a range of lattice sizes $\hat{L} \equiv L/a$.}
    \label{fig:var_ratios}
\end{figure}

Finally, we show in Fig.~\ref{fig:logz} the calculation of the log partition function ratio $\log(Z_e / Z_0)$ with the flow method and with standard approaches. In all cases, the uncertainties on the flow estimates are smaller than the plotted symbols.
We note that uncertainties on the baseline results are obtained by Gaussian error propagation applied to estimates of $Z_e/Z_0$, which results in underestimated uncertainties in cases where fluctuations are large and non-Gaussian; however, we observe convergence towards the flow results as the baseline statistics are increased. It is precisely the improvement in this estimate of the partition function ratio that results in more precise estimates of the correlator ratio and mass shift using the flow method. This effect is summarized in Fig.~\ref{fig:var_ratios}, which compares the factor improvement in the variance of correlator ratios $C_e/C_0$ as a function of separation $t/a$ versus the factor improvement in the variance of $\log(Z_e/Z_0)$ for both the linear and machine-learned flows.

\paragraph{Conclusions and outlook}---\,In this letter we considered a novel approach based on normalizing flows to determine unbiased all-orders QED effects in matrix elements in lattice field theory. The new framework was exemplified within the context of scalar QED in $N_d=2,3,4$ dimensions, and it was shown that the achievable precision outperforms brute-force stochastic approaches at fixed statistics. While the machine-learned flows must be trained at substantial computational cost, the learned model can be reused for evaluation on other lattice geometries. We showed that this volume transfer from a small training geometry to evaluation at a larger volume performs well. The new approach offers an alternative way to determine all-orders QED effects, which in the future could be applied to currently unaccessible matrix elements in QCD+QED, using well-established methods to apply normalizing flows to non-Abelian gauge theories and theories with fermions~\cite{Boyda:2020hsi,Albergo:2022qfi,Abbott:2022zhs}.
While an equivalent to linear flows is not straightforward to extend analytically, its effectiveness here suggests a machine-learning approach in a full dynamical theory may be able to realize similar gains.
In the long run, cost of training could be best amortized and the greatest impact achieved by developing more complex flow models which include parameter dependence of the underlying theory, allowing applications to many ensembles and potentially multiple regularization schemes.

\appendix

\begin{acknowledgments}
\paragraph{Acknowledgments}---\,The authors thank Maxwell T.~Hansen, Fernando Romero-L\'{o}pez, Antonin Portelli, and Phiala Shanahan for insightful discussions and for comments on the manuscript.
N.~H.-T.~is supported by the UK Research and Innovation, Science and Technology Facilities Council, grant number UKRI2426.
G.~K.~is supported by the Chancellor's Fellowship scheme in the University of Edinburgh.
The authors acknowledge the use of resources provided by the Isambard-AI National AI Research Resource (AIRR). Isambard-AI is operated by the University of Bristol and is funded by the UK Government’s Department for Science, Innovation and Technology (DSIT) via UK Research and Innovation; and the Science and Technology Facilities Council [ST/AIRR/I-A-I/1023]~\cite{mcintoshsmith2024isambardaileadershipclasssupercomputer}.
\end{acknowledgments}

\bibliography{refs.bib}

\providecommand{\noopsort}[1]{}\providecommand{\singleletter}[1]{#1}%
\begin{thebibliography}{65}%
\makeatletter
\providecommand \@ifxundefined [1]{%
 \@ifx{#1\undefined}
}%
\providecommand \@ifnum [1]{%
 \ifnum #1\expandafter \@firstoftwo
 \else \expandafter \@secondoftwo
 \fi
}%
\providecommand \@ifx [1]{%
 \ifx #1\expandafter \@firstoftwo
 \else \expandafter \@secondoftwo
 \fi
}%
\providecommand \natexlab [1]{#1}%
\providecommand \enquote  [1]{``#1''}%
\providecommand \bibnamefont  [1]{#1}%
\providecommand \bibfnamefont [1]{#1}%
\providecommand \citenamefont [1]{#1}%
\providecommand \href@noop [0]{\@secondoftwo}%
\providecommand \href [0]{\begingroup \@sanitize@url \@href}%
\providecommand \@href[1]{\@@startlink{#1}\@@href}%
\providecommand \@@href[1]{\endgroup#1\@@endlink}%
\providecommand \@sanitize@url [0]{\catcode `\\12\catcode `\$12\catcode
  `\&12\catcode `\#12\catcode `\^12\catcode `\_12\catcode `\%12\relax}%
\providecommand \@@startlink[1]{}%
\providecommand \@@endlink[0]{}%
\providecommand \url  [0]{\begingroup\@sanitize@url \@url }%
\providecommand \@url [1]{\endgroup\@href {#1}{\urlprefix }}%
\providecommand \urlprefix  [0]{URL }%
\providecommand \Eprint [0]{\href }%
\providecommand \doibase [0]{https://doi.org/}%
\providecommand \selectlanguage [0]{\@gobble}%
\providecommand \bibinfo  [0]{\@secondoftwo}%
\providecommand \bibfield  [0]{\@secondoftwo}%
\providecommand \translation [1]{[#1]}%
\providecommand \BibitemOpen [0]{}%
\providecommand \bibitemStop [0]{}%
\providecommand \bibitemNoStop [0]{.\EOS\space}%
\providecommand \EOS [0]{\spacefactor3000\relax}%
\providecommand \BibitemShut  [1]{\csname bibitem#1\endcsname}%
\let\auto@bib@innerbib\@empty
\bibitem [{\citenamefont {Aoki}\ \emph {et~al.}(2026)\citenamefont {Aoki} \emph
  {et~al.}}]{FlavourLatticeAveragingGroupFLAG:2024oxs}%
  \BibitemOpen
  \bibfield  {author} {\bibinfo {author} {\bibfnamefont {Y.}~\bibnamefont
  {Aoki}} \emph {et~al.} (\bibinfo {collaboration} {Flavour Lattice Averaging
  Group (FLAG)}),\ }\bibfield  {title} {\bibinfo {title} {{FLAG review 2024}},\
  }\href {https://doi.org/10.1103/nfzp-p5dn} {\bibfield  {journal} {\bibinfo
  {journal} {Phys. Rev. D}\ }\textbf {\bibinfo {volume} {113}},\ \bibinfo
  {pages} {014508} (\bibinfo {year} {2026})},\ \Eprint
  {https://arxiv.org/abs/2411.04268} {arXiv:2411.04268 [hep-lat]} \BibitemShut
  {NoStop}%
\bibitem [{\citenamefont {Borsanyi}\ \emph {et~al.}(2015)\citenamefont
  {Borsanyi} \emph {et~al.}}]{BMW:2014pzb}%
  \BibitemOpen
  \bibfield  {author} {\bibinfo {author} {\bibfnamefont {S.}~\bibnamefont
  {Borsanyi}} \emph {et~al.} (\bibinfo {collaboration} {BMW}),\ }\bibfield
  {title} {\bibinfo {title} {{Ab initio calculation of the neutron-proton mass
  difference}},\ }\href {https://doi.org/10.1126/science.1257050} {\bibfield
  {journal} {\bibinfo  {journal} {Science}\ }\textbf {\bibinfo {volume}
  {347}},\ \bibinfo {pages} {1452} (\bibinfo {year} {2015})},\ \Eprint
  {https://arxiv.org/abs/1406.4088} {arXiv:1406.4088 [hep-lat]} \BibitemShut
  {NoStop}%
\bibitem [{\citenamefont {de~Divitiis}\ \emph {et~al.}(2013)\citenamefont
  {de~Divitiis}, \citenamefont {Frezzotti}, \citenamefont {Lubicz},
  \citenamefont {Martinelli}, \citenamefont {Petronzio}, \citenamefont {Rossi},
  \citenamefont {Sanfilippo}, \citenamefont {Simula},\ and\ \citenamefont
  {Tantalo}}]{deDivitiis:2013xla}%
  \BibitemOpen
  \bibfield  {author} {\bibinfo {author} {\bibfnamefont {G.~M.}\ \bibnamefont
  {de~Divitiis}}, \bibinfo {author} {\bibfnamefont {R.}~\bibnamefont
  {Frezzotti}}, \bibinfo {author} {\bibfnamefont {V.}~\bibnamefont {Lubicz}},
  \bibinfo {author} {\bibfnamefont {G.}~\bibnamefont {Martinelli}}, \bibinfo
  {author} {\bibfnamefont {R.}~\bibnamefont {Petronzio}}, \bibinfo {author}
  {\bibfnamefont {G.~C.}\ \bibnamefont {Rossi}}, \bibinfo {author}
  {\bibfnamefont {F.}~\bibnamefont {Sanfilippo}}, \bibinfo {author}
  {\bibfnamefont {S.}~\bibnamefont {Simula}},\ and\ \bibinfo {author}
  {\bibfnamefont {N.}~\bibnamefont {Tantalo}} (\bibinfo {collaboration}
  {RM123}),\ }\bibfield  {title} {\bibinfo {title} {{Leading isospin breaking
  effects on the lattice}},\ }\href
  {https://doi.org/10.1103/PhysRevD.87.114505} {\bibfield  {journal} {\bibinfo
  {journal} {Phys. Rev. D}\ }\textbf {\bibinfo {volume} {87}},\ \bibinfo
  {pages} {114505} (\bibinfo {year} {2013})},\ \Eprint
  {https://arxiv.org/abs/1303.4896} {arXiv:1303.4896 [hep-lat]} \BibitemShut
  {NoStop}%
\bibitem [{\citenamefont {Clark}\ \emph {et~al.}(2022)\citenamefont {Clark},
  \citenamefont {Della~Morte}, \citenamefont {Hall}, \citenamefont {H{\"o}rz},
  \citenamefont {Nicholson}, \citenamefont {Shindler}, \citenamefont {Tsang},
  \citenamefont {Walker-Loud},\ and\ \citenamefont {Yan}}]{Clark:2022wjy}%
  \BibitemOpen
  \bibfield  {author} {\bibinfo {author} {\bibfnamefont {M.~A.}\ \bibnamefont
  {Clark}}, \bibinfo {author} {\bibfnamefont {M.}~\bibnamefont {Della~Morte}},
  \bibinfo {author} {\bibfnamefont {Z.}~\bibnamefont {Hall}}, \bibinfo {author}
  {\bibfnamefont {B.}~\bibnamefont {H{\"o}rz}}, \bibinfo {author}
  {\bibfnamefont {A.}~\bibnamefont {Nicholson}}, \bibinfo {author}
  {\bibfnamefont {A.}~\bibnamefont {Shindler}}, \bibinfo {author}
  {\bibfnamefont {J.~T.}\ \bibnamefont {Tsang}}, \bibinfo {author}
  {\bibfnamefont {A.}~\bibnamefont {Walker-Loud}},\ and\ \bibinfo {author}
  {\bibfnamefont {H.}~\bibnamefont {Yan}},\ }\bibfield  {title} {\bibinfo
  {title} {{QED with massive photons for precision physics: zero modes and
  first result for the hadron spectrum}},\ }\href
  {https://doi.org/10.22323/1.396.0281} {\bibfield  {journal} {\bibinfo
  {journal} {PoS}\ }\textbf {\bibinfo {volume} {LATTICE2021}},\ \bibinfo
  {pages} {281} (\bibinfo {year} {2022})},\ \Eprint
  {https://arxiv.org/abs/2201.03251} {arXiv:2201.03251 [hep-lat]} \BibitemShut
  {NoStop}%
\bibitem [{\citenamefont {Frezzotti}\ \emph {et~al.}(2022)\citenamefont
  {Frezzotti}, \citenamefont {Gagliardi}, \citenamefont {Lubicz}, \citenamefont
  {Martinelli}, \citenamefont {Sanfilippo},\ and\ \citenamefont
  {Simula}}]{Frezzotti:2022dwn}%
  \BibitemOpen
  \bibfield  {author} {\bibinfo {author} {\bibfnamefont {R.}~\bibnamefont
  {Frezzotti}}, \bibinfo {author} {\bibfnamefont {G.}~\bibnamefont
  {Gagliardi}}, \bibinfo {author} {\bibfnamefont {V.}~\bibnamefont {Lubicz}},
  \bibinfo {author} {\bibfnamefont {G.}~\bibnamefont {Martinelli}}, \bibinfo
  {author} {\bibfnamefont {F.}~\bibnamefont {Sanfilippo}},\ and\ \bibinfo
  {author} {\bibfnamefont {S.}~\bibnamefont {Simula}},\ }\bibfield  {title}
  {\bibinfo {title} {{Lattice calculation of the pion mass difference
  {\ensuremath{M_{\pi^+}}}-{\ensuremath{M_{\pi^0}}} at order
  O({\ensuremath{\alpha_{\rm em}}})}},\ }\href
  {https://doi.org/10.1103/PhysRevD.106.014502} {\bibfield  {journal} {\bibinfo
   {journal} {Phys. Rev. D}\ }\textbf {\bibinfo {volume} {106}},\ \bibinfo
  {pages} {014502} (\bibinfo {year} {2022})},\ \Eprint
  {https://arxiv.org/abs/2202.11970} {arXiv:2202.11970 [hep-lat]} \BibitemShut
  {NoStop}%
\bibitem [{\citenamefont {Bushnaq}\ \emph {et~al.}(2023)\citenamefont
  {Bushnaq}, \citenamefont {Campos}, \citenamefont {Catillo}, \citenamefont
  {Cotellucci}, \citenamefont {Dale}, \citenamefont {Fritzsch}, \citenamefont
  {L{\"u}cke}, \citenamefont {Krsti{\'c}~Marinkovi{\'c}}, \citenamefont
  {Patella},\ and\ \citenamefont {Tantalo}}]{RCstar:2022yjz}%
  \BibitemOpen
  \bibfield  {author} {\bibinfo {author} {\bibfnamefont {L.}~\bibnamefont
  {Bushnaq}}, \bibinfo {author} {\bibfnamefont {I.}~\bibnamefont {Campos}},
  \bibinfo {author} {\bibfnamefont {M.}~\bibnamefont {Catillo}}, \bibinfo
  {author} {\bibfnamefont {A.}~\bibnamefont {Cotellucci}}, \bibinfo {author}
  {\bibfnamefont {M.}~\bibnamefont {Dale}}, \bibinfo {author} {\bibfnamefont
  {P.}~\bibnamefont {Fritzsch}}, \bibinfo {author} {\bibfnamefont
  {J.}~\bibnamefont {L{\"u}cke}}, \bibinfo {author} {\bibfnamefont
  {M.}~\bibnamefont {Krsti{\'c}~Marinkovi{\'c}}}, \bibinfo {author}
  {\bibfnamefont {A.}~\bibnamefont {Patella}},\ and\ \bibinfo {author}
  {\bibfnamefont {N.}~\bibnamefont {Tantalo}} (\bibinfo {collaboration}
  {RCstar}),\ }\bibfield  {title} {\bibinfo {title} {{First results on QCD+QED
  with C$^{*}$ boundary conditions}},\ }\href
  {https://doi.org/10.1007/JHEP03(2023)012} {\bibfield  {journal} {\bibinfo
  {journal} {JHEP}\ }\textbf {\bibinfo {volume} {03}},\ \bibinfo {pages}
  {012}},\ \Eprint {https://arxiv.org/abs/2209.13183} {arXiv:2209.13183
  [hep-lat]} \BibitemShut {NoStop}%
\bibitem [{\citenamefont {Altherr}\ \emph
  {et~al.}(2026{\natexlab{a}})\citenamefont {Altherr}, \citenamefont {Campos},
  \citenamefont {Gruber}, \citenamefont {Harris}, \citenamefont {Margari},
  \citenamefont {Krsti{\'c}~Marinkovi{\'c}}, \citenamefont {Parato},
  \citenamefont {Patella}, \citenamefont {Rosso},\ and\ \citenamefont
  {Tavella}}]{Altherr:2026pux}%
  \BibitemOpen
  \bibfield  {author} {\bibinfo {author} {\bibfnamefont {A.}~\bibnamefont
  {Altherr}}, \bibinfo {author} {\bibfnamefont {I.}~\bibnamefont {Campos}},
  \bibinfo {author} {\bibfnamefont {R.}~\bibnamefont {Gruber}}, \bibinfo
  {author} {\bibfnamefont {T.}~\bibnamefont {Harris}}, \bibinfo {author}
  {\bibfnamefont {F.}~\bibnamefont {Margari}}, \bibinfo {author} {\bibfnamefont
  {M.}~\bibnamefont {Krsti{\'c}~Marinkovi{\'c}}}, \bibinfo {author}
  {\bibfnamefont {L.}~\bibnamefont {Parato}}, \bibinfo {author} {\bibfnamefont
  {A.}~\bibnamefont {Patella}}, \bibinfo {author} {\bibfnamefont
  {S.}~\bibnamefont {Rosso}},\ and\ \bibinfo {author} {\bibfnamefont
  {P.}~\bibnamefont {Tavella}},\ }\bibfield  {title} {\bibinfo {title} {{Baryon
  masses with C-periodic boundary conditions}},\ }\href@noop {} {\bibfield
  {journal} {\bibinfo  {journal} {arXiv}\ } (\bibinfo {year}
  {2026}{\natexlab{a}})},\ \Eprint {https://arxiv.org/abs/2602.23910}
  {arXiv:2602.23910 [hep-lat]} \BibitemShut {NoStop}%
\bibitem [{\citenamefont {Carrasco}\ \emph {et~al.}(2015)\citenamefont
  {Carrasco}, \citenamefont {Lubicz}, \citenamefont {Martinelli}, \citenamefont
  {Sachrajda}, \citenamefont {Tantalo}, \citenamefont {Tarantino},\ and\
  \citenamefont {Testa}}]{Carrasco:2015xwa}%
  \BibitemOpen
  \bibfield  {author} {\bibinfo {author} {\bibfnamefont {N.}~\bibnamefont
  {Carrasco}}, \bibinfo {author} {\bibfnamefont {V.}~\bibnamefont {Lubicz}},
  \bibinfo {author} {\bibfnamefont {G.}~\bibnamefont {Martinelli}}, \bibinfo
  {author} {\bibfnamefont {C.~T.}\ \bibnamefont {Sachrajda}}, \bibinfo {author}
  {\bibfnamefont {N.}~\bibnamefont {Tantalo}}, \bibinfo {author} {\bibfnamefont
  {C.}~\bibnamefont {Tarantino}},\ and\ \bibinfo {author} {\bibfnamefont
  {M.}~\bibnamefont {Testa}},\ }\bibfield  {title} {\bibinfo {title} {{QED
  Corrections to Hadronic Processes in Lattice QCD}},\ }\href
  {https://doi.org/10.1103/PhysRevD.91.074506} {\bibfield  {journal} {\bibinfo
  {journal} {Phys. Rev. D}\ }\textbf {\bibinfo {volume} {91}},\ \bibinfo
  {pages} {074506} (\bibinfo {year} {2015})},\ \Eprint
  {https://arxiv.org/abs/1502.00257} {arXiv:1502.00257 [hep-lat]} \BibitemShut
  {NoStop}%
\bibitem [{\citenamefont {Giusti}\ \emph {et~al.}(2018)\citenamefont {Giusti},
  \citenamefont {Lubicz}, \citenamefont {Martinelli}, \citenamefont
  {Sachrajda}, \citenamefont {Sanfilippo}, \citenamefont {Simula},
  \citenamefont {Tantalo},\ and\ \citenamefont {Tarantino}}]{Giusti:2017dwk}%
  \BibitemOpen
  \bibfield  {author} {\bibinfo {author} {\bibfnamefont {D.}~\bibnamefont
  {Giusti}}, \bibinfo {author} {\bibfnamefont {V.}~\bibnamefont {Lubicz}},
  \bibinfo {author} {\bibfnamefont {G.}~\bibnamefont {Martinelli}}, \bibinfo
  {author} {\bibfnamefont {C.~T.}\ \bibnamefont {Sachrajda}}, \bibinfo {author}
  {\bibfnamefont {F.}~\bibnamefont {Sanfilippo}}, \bibinfo {author}
  {\bibfnamefont {S.}~\bibnamefont {Simula}}, \bibinfo {author} {\bibfnamefont
  {N.}~\bibnamefont {Tantalo}},\ and\ \bibinfo {author} {\bibfnamefont
  {C.}~\bibnamefont {Tarantino}},\ }\bibfield  {title} {\bibinfo {title}
  {{First lattice calculation of the QED corrections to leptonic decay
  rates}},\ }\href {https://doi.org/10.1103/PhysRevLett.120.072001} {\bibfield
  {journal} {\bibinfo  {journal} {Phys. Rev. Lett.}\ }\textbf {\bibinfo
  {volume} {120}},\ \bibinfo {pages} {072001} (\bibinfo {year} {2018})},\
  \Eprint {https://arxiv.org/abs/1711.06537} {arXiv:1711.06537 [hep-lat]}
  \BibitemShut {NoStop}%
\bibitem [{\citenamefont {Di~Carlo}\ \emph {et~al.}(2019)\citenamefont
  {Di~Carlo}, \citenamefont {Giusti}, \citenamefont {Lubicz}, \citenamefont
  {Martinelli}, \citenamefont {Sachrajda}, \citenamefont {Sanfilippo},
  \citenamefont {Simula},\ and\ \citenamefont {Tantalo}}]{DiCarlo:2019thl}%
  \BibitemOpen
  \bibfield  {author} {\bibinfo {author} {\bibfnamefont {M.}~\bibnamefont
  {Di~Carlo}}, \bibinfo {author} {\bibfnamefont {D.}~\bibnamefont {Giusti}},
  \bibinfo {author} {\bibfnamefont {V.}~\bibnamefont {Lubicz}}, \bibinfo
  {author} {\bibfnamefont {G.}~\bibnamefont {Martinelli}}, \bibinfo {author}
  {\bibfnamefont {C.~T.}\ \bibnamefont {Sachrajda}}, \bibinfo {author}
  {\bibfnamefont {F.}~\bibnamefont {Sanfilippo}}, \bibinfo {author}
  {\bibfnamefont {S.}~\bibnamefont {Simula}},\ and\ \bibinfo {author}
  {\bibfnamefont {N.}~\bibnamefont {Tantalo}},\ }\bibfield  {title} {\bibinfo
  {title} {{Light-meson leptonic decay rates in lattice QCD+QED}},\ }\href
  {https://doi.org/10.1103/PhysRevD.100.034514} {\bibfield  {journal} {\bibinfo
   {journal} {Phys. Rev. D}\ }\textbf {\bibinfo {volume} {100}},\ \bibinfo
  {pages} {034514} (\bibinfo {year} {2019})},\ \Eprint
  {https://arxiv.org/abs/1904.08731} {arXiv:1904.08731 [hep-lat]} \BibitemShut
  {NoStop}%
\bibitem [{\citenamefont {Boyle}\ \emph {et~al.}(2023)\citenamefont {Boyle}
  \emph {et~al.}}]{Boyle:2022lsi}%
  \BibitemOpen
  \bibfield  {author} {\bibinfo {author} {\bibfnamefont {P.}~\bibnamefont
  {Boyle}} \emph {et~al.},\ }\bibfield  {title} {\bibinfo {title}
  {{Isospin-breaking corrections to light-meson leptonic decays from lattice
  simulations at physical quark masses}},\ }\href
  {https://doi.org/10.1007/JHEP02(2023)242} {\bibfield  {journal} {\bibinfo
  {journal} {JHEP}\ }\textbf {\bibinfo {volume} {02}},\ \bibinfo {pages}
  {242}},\ \Eprint {https://arxiv.org/abs/2211.12865} {arXiv:2211.12865
  [hep-lat]} \BibitemShut {NoStop}%
\bibitem [{\citenamefont {Christ}\ \emph {et~al.}(2023)\citenamefont {Christ},
  \citenamefont {Feng}, \citenamefont {Jin}, \citenamefont {Sachrajda},\ and\
  \citenamefont {Wang}}]{Christ:2023lcc}%
  \BibitemOpen
  \bibfield  {author} {\bibinfo {author} {\bibfnamefont {N.~H.}\ \bibnamefont
  {Christ}}, \bibinfo {author} {\bibfnamefont {X.}~\bibnamefont {Feng}},
  \bibinfo {author} {\bibfnamefont {L.-C.}\ \bibnamefont {Jin}}, \bibinfo
  {author} {\bibfnamefont {C.~T.}\ \bibnamefont {Sachrajda}},\ and\ \bibinfo
  {author} {\bibfnamefont {T.}~\bibnamefont {Wang}},\ }\bibfield  {title}
  {\bibinfo {title} {{Radiative corrections to leptonic decays using
  infinite-volume reconstruction}},\ }\href
  {https://doi.org/10.1103/PhysRevD.108.014501} {\bibfield  {journal} {\bibinfo
   {journal} {Phys. Rev. D}\ }\textbf {\bibinfo {volume} {108}},\ \bibinfo
  {pages} {014501} (\bibinfo {year} {2023})},\ \Eprint
  {https://arxiv.org/abs/2304.08026} {arXiv:2304.08026 [hep-lat]} \BibitemShut
  {NoStop}%
\bibitem [{\citenamefont {Giusti}\ \emph
  {et~al.}(2017{\natexlab{a}})\citenamefont {Giusti}, \citenamefont {Lubicz},
  \citenamefont {Martinelli}, \citenamefont {Sanfilippo},\ and\ \citenamefont
  {Simula}}]{Giusti:2017jof}%
  \BibitemOpen
  \bibfield  {author} {\bibinfo {author} {\bibfnamefont {D.}~\bibnamefont
  {Giusti}}, \bibinfo {author} {\bibfnamefont {V.}~\bibnamefont {Lubicz}},
  \bibinfo {author} {\bibfnamefont {G.}~\bibnamefont {Martinelli}}, \bibinfo
  {author} {\bibfnamefont {F.}~\bibnamefont {Sanfilippo}},\ and\ \bibinfo
  {author} {\bibfnamefont {S.}~\bibnamefont {Simula}},\ }\bibfield  {title}
  {\bibinfo {title} {{Strange and charm HVP contributions to the muon ($g - 2)$
  including QED corrections with twisted-mass fermions}},\ }\href
  {https://doi.org/10.1007/JHEP10(2017)157} {\bibfield  {journal} {\bibinfo
  {journal} {JHEP}\ }\textbf {\bibinfo {volume} {10}},\ \bibinfo {pages}
  {157}},\ \Eprint {https://arxiv.org/abs/1707.03019} {arXiv:1707.03019
  [hep-lat]} \BibitemShut {NoStop}%
\bibitem [{\citenamefont {Blum}\ \emph {et~al.}(2018)\citenamefont {Blum},
  \citenamefont {Boyle}, \citenamefont {G{\"u}lpers}, \citenamefont {Izubuchi},
  \citenamefont {Jin}, \citenamefont {Jung}, \citenamefont {J{\"u}ttner},
  \citenamefont {Lehner}, \citenamefont {Portelli},\ and\ \citenamefont
  {Tsang}}]{RBC:2018dos}%
  \BibitemOpen
  \bibfield  {author} {\bibinfo {author} {\bibfnamefont {T.}~\bibnamefont
  {Blum}}, \bibinfo {author} {\bibfnamefont {P.~A.}\ \bibnamefont {Boyle}},
  \bibinfo {author} {\bibfnamefont {V.}~\bibnamefont {G{\"u}lpers}}, \bibinfo
  {author} {\bibfnamefont {T.}~\bibnamefont {Izubuchi}}, \bibinfo {author}
  {\bibfnamefont {L.}~\bibnamefont {Jin}}, \bibinfo {author} {\bibfnamefont
  {C.}~\bibnamefont {Jung}}, \bibinfo {author} {\bibfnamefont {A.}~\bibnamefont
  {J{\"u}ttner}}, \bibinfo {author} {\bibfnamefont {C.}~\bibnamefont {Lehner}},
  \bibinfo {author} {\bibfnamefont {A.}~\bibnamefont {Portelli}},\ and\
  \bibinfo {author} {\bibfnamefont {J.~T.}\ \bibnamefont {Tsang}} (\bibinfo
  {collaboration} {RBC, UKQCD}),\ }\bibfield  {title} {\bibinfo {title}
  {{Calculation of the hadronic vacuum polarization contribution to the muon
  anomalous magnetic moment}},\ }\href
  {https://doi.org/10.1103/PhysRevLett.121.022003} {\bibfield  {journal}
  {\bibinfo  {journal} {Phys. Rev. Lett.}\ }\textbf {\bibinfo {volume} {121}},\
  \bibinfo {pages} {022003} (\bibinfo {year} {2018})},\ \Eprint
  {https://arxiv.org/abs/1801.07224} {arXiv:1801.07224 [hep-lat]} \BibitemShut
  {NoStop}%
\bibitem [{\citenamefont {Giusti}\ \emph
  {et~al.}(2017{\natexlab{b}})\citenamefont {Giusti}, \citenamefont {Lubicz},
  \citenamefont {Tarantino}, \citenamefont {Martinelli}, \citenamefont
  {Sanfilippo}, \citenamefont {Simula},\ and\ \citenamefont
  {Tantalo}}]{Giusti:2017dmp}%
  \BibitemOpen
  \bibfield  {author} {\bibinfo {author} {\bibfnamefont {D.}~\bibnamefont
  {Giusti}}, \bibinfo {author} {\bibfnamefont {V.}~\bibnamefont {Lubicz}},
  \bibinfo {author} {\bibfnamefont {C.}~\bibnamefont {Tarantino}}, \bibinfo
  {author} {\bibfnamefont {G.}~\bibnamefont {Martinelli}}, \bibinfo {author}
  {\bibfnamefont {F.}~\bibnamefont {Sanfilippo}}, \bibinfo {author}
  {\bibfnamefont {S.}~\bibnamefont {Simula}},\ and\ \bibinfo {author}
  {\bibfnamefont {N.}~\bibnamefont {Tantalo}},\ }\bibfield  {title} {\bibinfo
  {title} {{Leading isospin-breaking corrections to pion, kaon and
  charmed-meson masses with Twisted-Mass fermions}},\ }\href
  {https://doi.org/10.1103/PhysRevD.95.114504} {\bibfield  {journal} {\bibinfo
  {journal} {Phys. Rev. D}\ }\textbf {\bibinfo {volume} {95}},\ \bibinfo
  {pages} {114504} (\bibinfo {year} {2017}{\natexlab{b}})},\ \Eprint
  {https://arxiv.org/abs/1704.06561} {arXiv:1704.06561 [hep-lat]} \BibitemShut
  {NoStop}%
\bibitem [{\citenamefont {Giusti}\ \emph {et~al.}(2019)\citenamefont {Giusti},
  \citenamefont {Lubicz}, \citenamefont {Martinelli}, \citenamefont
  {Sanfilippo},\ and\ \citenamefont {Simula}}]{Giusti:2019xct}%
  \BibitemOpen
  \bibfield  {author} {\bibinfo {author} {\bibfnamefont {D.}~\bibnamefont
  {Giusti}}, \bibinfo {author} {\bibfnamefont {V.}~\bibnamefont {Lubicz}},
  \bibinfo {author} {\bibfnamefont {G.}~\bibnamefont {Martinelli}}, \bibinfo
  {author} {\bibfnamefont {F.}~\bibnamefont {Sanfilippo}},\ and\ \bibinfo
  {author} {\bibfnamefont {S.}~\bibnamefont {Simula}},\ }\bibfield  {title}
  {\bibinfo {title} {{Electromagnetic and strong isospin-breaking corrections
  to the muon $g - 2$ from Lattice QCD+QED}},\ }\href
  {https://doi.org/10.1103/PhysRevD.99.114502} {\bibfield  {journal} {\bibinfo
  {journal} {Phys. Rev. D}\ }\textbf {\bibinfo {volume} {99}},\ \bibinfo
  {pages} {114502} (\bibinfo {year} {2019})},\ \Eprint
  {https://arxiv.org/abs/1901.10462} {arXiv:1901.10462 [hep-lat]} \BibitemShut
  {NoStop}%
\bibitem [{\citenamefont {Aoyama}\ \emph {et~al.}(2020)\citenamefont {Aoyama}
  \emph {et~al.}}]{Aoyama:2020ynm}%
  \BibitemOpen
  \bibfield  {author} {\bibinfo {author} {\bibfnamefont {T.}~\bibnamefont
  {Aoyama}} \emph {et~al.},\ }\bibfield  {title} {\bibinfo {title} {{The
  anomalous magnetic moment of the muon in the Standard Model}},\ }\href
  {https://doi.org/10.1016/j.physrep.2020.07.006} {\bibfield  {journal}
  {\bibinfo  {journal} {Phys. Rept.}\ }\textbf {\bibinfo {volume} {887}},\
  \bibinfo {pages} {1} (\bibinfo {year} {2020})},\ \Eprint
  {https://arxiv.org/abs/2006.04822} {arXiv:2006.04822 [hep-ph]} \BibitemShut
  {NoStop}%
\bibitem [{\citenamefont {Borsanyi}\ \emph {et~al.}(2021)\citenamefont
  {Borsanyi} \emph {et~al.}}]{Borsanyi:2020mff}%
  \BibitemOpen
  \bibfield  {author} {\bibinfo {author} {\bibfnamefont {S.}~\bibnamefont
  {Borsanyi}} \emph {et~al.},\ }\bibfield  {title} {\bibinfo {title} {{Leading
  hadronic contribution to the muon magnetic moment from lattice QCD}},\ }\href
  {https://doi.org/10.1038/s41586-021-03418-1} {\bibfield  {journal} {\bibinfo
  {journal} {Nature}\ }\textbf {\bibinfo {volume} {593}},\ \bibinfo {pages}
  {51} (\bibinfo {year} {2021})},\ \Eprint {https://arxiv.org/abs/2002.12347}
  {arXiv:2002.12347 [hep-lat]} \BibitemShut {NoStop}%
\bibitem [{\citenamefont {Biloshytskyi}\ \emph {et~al.}(2023)\citenamefont
  {Biloshytskyi}, \citenamefont {Chao}, \citenamefont {G\'erardin},
  \citenamefont {Green}, \citenamefont {Hagelstein}, \citenamefont {Meyer},
  \citenamefont {Parrino},\ and\ \citenamefont
  {Pascalutsa}}]{Biloshytskyi:2022ets}%
  \BibitemOpen
  \bibfield  {author} {\bibinfo {author} {\bibfnamefont {V.}~\bibnamefont
  {Biloshytskyi}}, \bibinfo {author} {\bibfnamefont {E.-H.}\ \bibnamefont
  {Chao}}, \bibinfo {author} {\bibfnamefont {A.}~\bibnamefont {G\'erardin}},
  \bibinfo {author} {\bibfnamefont {J.~R.}\ \bibnamefont {Green}}, \bibinfo
  {author} {\bibfnamefont {F.}~\bibnamefont {Hagelstein}}, \bibinfo {author}
  {\bibfnamefont {H.~B.}\ \bibnamefont {Meyer}}, \bibinfo {author}
  {\bibfnamefont {J.}~\bibnamefont {Parrino}},\ and\ \bibinfo {author}
  {\bibfnamefont {V.}~\bibnamefont {Pascalutsa}},\ }\bibfield  {title}
  {\bibinfo {title} {{Forward light-by-light scattering and electromagnetic
  correction to hadronic vacuum polarization}},\ }\href
  {https://doi.org/10.1007/JHEP03(2023)194} {\bibfield  {journal} {\bibinfo
  {journal} {JHEP}\ }\textbf {\bibinfo {volume} {03}},\ \bibinfo {pages}
  {194}},\ \Eprint {https://arxiv.org/abs/2209.02149} {arXiv:2209.02149
  [hep-lat]} \BibitemShut {NoStop}%
\bibitem [{\citenamefont {Chao}\ \emph {et~al.}(2024)\citenamefont {Chao},
  \citenamefont {Meyer},\ and\ \citenamefont {Parrino}}]{Chao:2023lxw}%
  \BibitemOpen
  \bibfield  {author} {\bibinfo {author} {\bibfnamefont {E.-H.}\ \bibnamefont
  {Chao}}, \bibinfo {author} {\bibfnamefont {H.~B.}\ \bibnamefont {Meyer}},\
  and\ \bibinfo {author} {\bibfnamefont {J.}~\bibnamefont {Parrino}},\
  }\bibfield  {title} {\bibinfo {title} {{Coordinate-space calculation of QED
  corrections to the hadronic vacuum polarization contribution to
  $(g-2)_\mu$}},\ }\href {https://doi.org/10.22323/1.453.0256} {\bibfield
  {journal} {\bibinfo  {journal} {PoS}\ }\textbf {\bibinfo {volume}
  {LATTICE2023}},\ \bibinfo {pages} {256} (\bibinfo {year} {2024})},\ \Eprint
  {https://arxiv.org/abs/2310.20556} {arXiv:2310.20556 [hep-lat]} \BibitemShut
  {NoStop}%
\bibitem [{\citenamefont {Boccaletti}\ \emph {et~al.}(2026)\citenamefont
  {Boccaletti} \emph {et~al.}}]{Boccaletti:2024guq}%
  \BibitemOpen
  \bibfield  {author} {\bibinfo {author} {\bibfnamefont {A.}~\bibnamefont
  {Boccaletti}} \emph {et~al.},\ }\bibfield  {title} {\bibinfo {title} {{Hybrid
  calculation of hadronic vacuum polarization in muon g {\ensuremath{-}} 2 to
  0.48{\%}}},\ }\href {https://doi.org/10.1038/s41586-026-10449-z} {\bibfield
  {journal} {\bibinfo  {journal} {Nature}\ }\textbf {\bibinfo {volume} {653}},\
  \bibinfo {pages} {373} (\bibinfo {year} {2026})},\ \Eprint
  {https://arxiv.org/abs/2407.10913} {arXiv:2407.10913 [hep-lat]} \BibitemShut
  {NoStop}%
\bibitem [{\citenamefont {Djukanovic}\ \emph {et~al.}(2025)\citenamefont
  {Djukanovic}, \citenamefont {von Hippel}, \citenamefont {Kuberski},
  \citenamefont {Meyer}, \citenamefont {Miller}, \citenamefont {Ottnad},
  \citenamefont {Parrino}, \citenamefont {Risch},\ and\ \citenamefont
  {Wittig}}]{Djukanovic:2024cmq}%
  \BibitemOpen
  \bibfield  {author} {\bibinfo {author} {\bibfnamefont {D.}~\bibnamefont
  {Djukanovic}}, \bibinfo {author} {\bibfnamefont {G.}~\bibnamefont {von
  Hippel}}, \bibinfo {author} {\bibfnamefont {S.}~\bibnamefont {Kuberski}},
  \bibinfo {author} {\bibfnamefont {H.~B.}\ \bibnamefont {Meyer}}, \bibinfo
  {author} {\bibfnamefont {N.}~\bibnamefont {Miller}}, \bibinfo {author}
  {\bibfnamefont {K.}~\bibnamefont {Ottnad}}, \bibinfo {author} {\bibfnamefont
  {J.}~\bibnamefont {Parrino}}, \bibinfo {author} {\bibfnamefont
  {A.}~\bibnamefont {Risch}},\ and\ \bibinfo {author} {\bibfnamefont
  {H.}~\bibnamefont {Wittig}},\ }\bibfield  {title} {\bibinfo {title} {{The
  hadronic vacuum polarization contribution to the muon g {\ensuremath{-}} 2 at
  long distances}},\ }\href {https://doi.org/10.1007/JHEP04(2025)098}
  {\bibfield  {journal} {\bibinfo  {journal} {JHEP}\ }\textbf {\bibinfo
  {volume} {04}},\ \bibinfo {pages} {098}},\ \Eprint
  {https://arxiv.org/abs/2411.07969} {arXiv:2411.07969 [hep-lat]} \BibitemShut
  {NoStop}%
\bibitem [{\citenamefont {Parrino}\ \emph {et~al.}(2025)\citenamefont
  {Parrino}, \citenamefont {Biloshytskyi}, \citenamefont {Chao}, \citenamefont
  {Meyer},\ and\ \citenamefont {Pascalutsa}}]{Parrino:2025afq}%
  \BibitemOpen
  \bibfield  {author} {\bibinfo {author} {\bibfnamefont {J.}~\bibnamefont
  {Parrino}}, \bibinfo {author} {\bibfnamefont {V.}~\bibnamefont
  {Biloshytskyi}}, \bibinfo {author} {\bibfnamefont {E.-H.}\ \bibnamefont
  {Chao}}, \bibinfo {author} {\bibfnamefont {H.~B.}\ \bibnamefont {Meyer}},\
  and\ \bibinfo {author} {\bibfnamefont {V.}~\bibnamefont {Pascalutsa}},\
  }\bibfield  {title} {\bibinfo {title} {{Computing the UV-finite
  electromagnetic corrections to the hadronic vacuum polarization in the muon
  (g {\ensuremath{-}} 2) from lattice QCD}},\ }\href
  {https://doi.org/10.1007/JHEP07(2025)201} {\bibfield  {journal} {\bibinfo
  {journal} {JHEP}\ }\textbf {\bibinfo {volume} {07}},\ \bibinfo {pages}
  {201}},\ \Eprint {https://arxiv.org/abs/2501.03192} {arXiv:2501.03192
  [hep-lat]} \BibitemShut {NoStop}%
\bibitem [{\citenamefont {Altherr}\ \emph {et~al.}(2025)\citenamefont {Altherr}
  \emph {et~al.}}]{RC:2025zoa}%
  \BibitemOpen
  \bibfield  {author} {\bibinfo {author} {\bibfnamefont {A.}~\bibnamefont
  {Altherr}} \emph {et~al.} (\bibinfo {collaboration}
  {RC{\ensuremath{\star}}}),\ }\bibfield  {title} {\bibinfo {title} {{Comparing
  QCD+QED via full simulation versus the RM123 method: U-spin window
  contribution to ${a}_{\mu }^{\text{HVP}}$}},\ }\href
  {https://doi.org/10.1007/JHEP10(2025)158} {\bibfield  {journal} {\bibinfo
  {journal} {JHEP}\ }\textbf {\bibinfo {volume} {10}},\ \bibinfo {pages}
  {158}},\ \Eprint {https://arxiv.org/abs/2506.19770} {arXiv:2506.19770
  [hep-lat]} \BibitemShut {NoStop}%
\bibitem [{\citenamefont {Aliberti}\ \emph {et~al.}(2025)\citenamefont
  {Aliberti} \emph {et~al.}}]{Aliberti:2025beg}%
  \BibitemOpen
  \bibfield  {author} {\bibinfo {author} {\bibfnamefont {R.}~\bibnamefont
  {Aliberti}} \emph {et~al.},\ }\bibfield  {title} {\bibinfo {title} {{The
  anomalous magnetic moment of the muon in the Standard Model: an update}},\
  }\href {https://doi.org/10.1016/j.physrep.2025.08.002} {\bibfield  {journal}
  {\bibinfo  {journal} {Phys. Rept.}\ }\textbf {\bibinfo {volume} {1143}},\
  \bibinfo {pages} {1} (\bibinfo {year} {2025})},\ \Eprint
  {https://arxiv.org/abs/2505.21476} {arXiv:2505.21476 [hep-ph]} \BibitemShut
  {NoStop}%
\bibitem [{\citenamefont {Bruno}\ \emph {et~al.}(2026)\citenamefont {Bruno},
  \citenamefont {G{\"u}lpers}, \citenamefont {Hermansson-Truedsson},
  \citenamefont {Lehner}, \citenamefont {Parrino},\ and\ \citenamefont
  {Tsang}}]{Bruno:2026kqf}%
  \BibitemOpen
  \bibfield  {author} {\bibinfo {author} {\bibfnamefont {M.}~\bibnamefont
  {Bruno}}, \bibinfo {author} {\bibfnamefont {V.}~\bibnamefont {G{\"u}lpers}},
  \bibinfo {author} {\bibfnamefont {N.}~\bibnamefont {Hermansson-Truedsson}},
  \bibinfo {author} {\bibfnamefont {C.}~\bibnamefont {Lehner}}, \bibinfo
  {author} {\bibfnamefont {J.}~\bibnamefont {Parrino}},\ and\ \bibinfo {author}
  {\bibfnamefont {J.~T.}\ \bibnamefont {Tsang}},\ }\bibfield  {title} {\bibinfo
  {title} {{Isospin breaking corrections to the hadronic vacuum polarization
  with stochastic coordinate sampling}},\ }in\ \href@noop {} {\emph {\bibinfo
  {booktitle} {{42th International Symposium on Lattice Field Theory}}}}\
  (\bibinfo {year} {2026})\ \Eprint {https://arxiv.org/abs/2602.24132}
  {arXiv:2602.24132 [hep-lat]} \BibitemShut {NoStop}%
\bibitem [{\citenamefont {Tomalak}\ and\ \citenamefont
  {Yang}(2026)}]{Tomalak:2026wks}%
  \BibitemOpen
  \bibfield  {author} {\bibinfo {author} {\bibfnamefont {O.}~\bibnamefont
  {Tomalak}}\ and\ \bibinfo {author} {\bibfnamefont {Y.-B.}\ \bibnamefont
  {Yang}},\ }\bibfield  {title} {\bibinfo {title} {{Radiative corrections to
  the nucleon isovector $g_V$ and $g_A$}},\ }\href@noop {} {\bibfield
  {journal} {\bibinfo  {journal} {Universe}\ }\textbf {\bibinfo {volume}
  {12}},\ \bibinfo {pages} {109} (\bibinfo {year} {2026})},\ \Eprint
  {https://arxiv.org/abs/2603.08596} {arXiv:2603.08596 [hep-ph]} \BibitemShut
  {NoStop}%
\bibitem [{\citenamefont {Cirigliano}\ \emph {et~al.}(2025)\citenamefont
  {Cirigliano}, \citenamefont {Dekens}, \citenamefont {Mereghetti},\ and\
  \citenamefont {Tomalak}}]{Cirigliano:2024nfi}%
  \BibitemOpen
  \bibfield  {author} {\bibinfo {author} {\bibfnamefont {V.}~\bibnamefont
  {Cirigliano}}, \bibinfo {author} {\bibfnamefont {W.}~\bibnamefont {Dekens}},
  \bibinfo {author} {\bibfnamefont {E.}~\bibnamefont {Mereghetti}},\ and\
  \bibinfo {author} {\bibfnamefont {O.}~\bibnamefont {Tomalak}},\ }\bibfield
  {title} {\bibinfo {title} {{Effective field theory for radiative corrections
  to charged-current processes. II. Axial-vector coupling}},\ }\href
  {https://doi.org/10.1103/PhysRevD.111.053005} {\bibfield  {journal} {\bibinfo
   {journal} {Phys. Rev. D}\ }\textbf {\bibinfo {volume} {111}},\ \bibinfo
  {pages} {053005} (\bibinfo {year} {2025})},\ \Eprint
  {https://arxiv.org/abs/2410.21404} {arXiv:2410.21404 [nucl-th]} \BibitemShut
  {NoStop}%
\bibitem [{\citenamefont {Gorchtein}\ and\ \citenamefont
  {Seng}(2023)}]{Gorchtein:2023srs}%
  \BibitemOpen
  \bibfield  {author} {\bibinfo {author} {\bibfnamefont {M.}~\bibnamefont
  {Gorchtein}}\ and\ \bibinfo {author} {\bibfnamefont {C.-Y.}\ \bibnamefont
  {Seng}},\ }\bibfield  {title} {\bibinfo {title} {{The Standard Model Theory
  of Neutron Beta Decay}},\ }\href {https://doi.org/10.3390/universe9090422}
  {\bibfield  {journal} {\bibinfo  {journal} {Universe}\ }\textbf {\bibinfo
  {volume} {9}},\ \bibinfo {pages} {422} (\bibinfo {year} {2023})},\ \Eprint
  {https://arxiv.org/abs/2307.01145} {arXiv:2307.01145 [hep-ph]} \BibitemShut
  {NoStop}%
\bibitem [{\citenamefont {Tabak}\ and\ \citenamefont
  {Vanden-Eijnden}(2010)}]{tabak2010}%
  \BibitemOpen
  \bibfield  {author} {\bibinfo {author} {\bibfnamefont {E.~G.}\ \bibnamefont
  {Tabak}}\ and\ \bibinfo {author} {\bibfnamefont {E.}~\bibnamefont
  {Vanden-Eijnden}},\ }\bibfield  {title} {\bibinfo {title} {Density estimation
  by dual ascent of the log-likelihood},\ }\href
  {https://doi.org/10.4310/CMS.2010.v8.n1.a11} {\bibfield  {journal} {\bibinfo
  {journal} {Commun. Math. Sci.}\ }\textbf {\bibinfo {volume} {8}},\ \bibinfo
  {pages} {217} (\bibinfo {year} {2010})}\BibitemShut {NoStop}%
\bibitem [{\citenamefont {Tabak}\ and\ \citenamefont
  {Turner}(2013)}]{tabak2013}%
  \BibitemOpen
  \bibfield  {author} {\bibinfo {author} {\bibfnamefont {E.~G.}\ \bibnamefont
  {Tabak}}\ and\ \bibinfo {author} {\bibfnamefont {C.~V.}\ \bibnamefont
  {Turner}},\ }\bibfield  {title} {\bibinfo {title} {A family of nonparametric
  density estimation algorithms},\ }\href
  {https://doi.org/https://doi.org/10.1002/cpa.21423} {\bibfield  {journal}
  {\bibinfo  {journal} {Communications on Pure and Applied Mathematics}\
  }\textbf {\bibinfo {volume} {66}},\ \bibinfo {pages} {145} (\bibinfo {year}
  {2013})}\BibitemShut {NoStop}%
\bibitem [{\citenamefont {Rezende}\ and\ \citenamefont
  {Mohamed}(2015)}]{rezende2015variational}%
  \BibitemOpen
  \bibfield  {author} {\bibinfo {author} {\bibfnamefont {D.}~\bibnamefont
  {Rezende}}\ and\ \bibinfo {author} {\bibfnamefont {S.}~\bibnamefont
  {Mohamed}},\ }\bibfield  {title} {\bibinfo {title} {Variational inference
  with normalizing flows},\ }in\ \href@noop {} {\emph {\bibinfo {booktitle}
  {International conference on machine learning}}}\ (\bibinfo {organization}
  {PMLR},\ \bibinfo {year} {2015})\ pp.\ \bibinfo {pages} {1530--1538},\
  \Eprint {https://arxiv.org/abs/1505.05770} {arXiv:1505.05770 [stat.ML]}
  \BibitemShut {NoStop}%
\bibitem [{\citenamefont {Albergo}\ \emph {et~al.}(2019)\citenamefont
  {Albergo}, \citenamefont {Kanwar},\ and\ \citenamefont
  {Shanahan}}]{Albergo:2019eim}%
  \BibitemOpen
  \bibfield  {author} {\bibinfo {author} {\bibfnamefont {M.~S.}\ \bibnamefont
  {Albergo}}, \bibinfo {author} {\bibfnamefont {G.}~\bibnamefont {Kanwar}},\
  and\ \bibinfo {author} {\bibfnamefont {P.~E.}\ \bibnamefont {Shanahan}},\
  }\bibfield  {title} {\bibinfo {title} {{Flow-based generative models for
  Markov chain Monte Carlo in lattice field theory}},\ }\href
  {https://doi.org/10.1103/PhysRevD.100.034515} {\bibfield  {journal} {\bibinfo
   {journal} {Phys. Rev. D}\ }\textbf {\bibinfo {volume} {100}},\ \bibinfo
  {pages} {034515} (\bibinfo {year} {2019})},\ \Eprint
  {https://arxiv.org/abs/1904.12072} {arXiv:1904.12072 [hep-lat]} \BibitemShut
  {NoStop}%
\bibitem [{\citenamefont {Abbott}\ \emph {et~al.}(2024)\citenamefont {Abbott},
  \citenamefont {Botev}, \citenamefont {Boyda}, \citenamefont {Hackett},
  \citenamefont {Kanwar}, \citenamefont {Racani{\`e}re}, \citenamefont
  {Rezende}, \citenamefont {Romero-L{\'o}pez}, \citenamefont {Shanahan},\ and\
  \citenamefont {Urban}}]{Abbott:2024kfc}%
  \BibitemOpen
  \bibfield  {author} {\bibinfo {author} {\bibfnamefont {R.}~\bibnamefont
  {Abbott}}, \bibinfo {author} {\bibfnamefont {A.}~\bibnamefont {Botev}},
  \bibinfo {author} {\bibfnamefont {D.}~\bibnamefont {Boyda}}, \bibinfo
  {author} {\bibfnamefont {D.~C.}\ \bibnamefont {Hackett}}, \bibinfo {author}
  {\bibfnamefont {G.}~\bibnamefont {Kanwar}}, \bibinfo {author} {\bibfnamefont
  {S.}~\bibnamefont {Racani{\`e}re}}, \bibinfo {author} {\bibfnamefont {D.~J.}\
  \bibnamefont {Rezende}}, \bibinfo {author} {\bibfnamefont {F.}~\bibnamefont
  {Romero-L{\'o}pez}}, \bibinfo {author} {\bibfnamefont {P.~E.}\ \bibnamefont
  {Shanahan}},\ and\ \bibinfo {author} {\bibfnamefont {J.~M.}\ \bibnamefont
  {Urban}},\ }\bibfield  {title} {\bibinfo {title} {{Applications of flow
  models to the generation of correlated lattice QCD ensembles}},\ }\href
  {https://doi.org/10.1103/PhysRevD.109.094514} {\bibfield  {journal} {\bibinfo
   {journal} {Phys. Rev. D}\ }\textbf {\bibinfo {volume} {109}},\ \bibinfo
  {pages} {094514} (\bibinfo {year} {2024})},\ \Eprint
  {https://arxiv.org/abs/2401.10874} {arXiv:2401.10874 [hep-lat]} \BibitemShut
  {NoStop}%
\bibitem [{\citenamefont {Abbott}\ \emph {et~al.}(2026)\citenamefont {Abbott},
  \citenamefont {Boyda}, \citenamefont {Fu}, \citenamefont {Hackett},
  \citenamefont {Kanwar}, \citenamefont {Romero-L{\'o}pez}, \citenamefont
  {Shanahan},\ and\ \citenamefont {Urban}}]{Abbott:2026ylv}%
  \BibitemOpen
  \bibfield  {author} {\bibinfo {author} {\bibfnamefont {R.}~\bibnamefont
  {Abbott}}, \bibinfo {author} {\bibfnamefont {D.}~\bibnamefont {Boyda}},
  \bibinfo {author} {\bibfnamefont {Y.}~\bibnamefont {Fu}}, \bibinfo {author}
  {\bibfnamefont {D.~C.}\ \bibnamefont {Hackett}}, \bibinfo {author}
  {\bibfnamefont {G.}~\bibnamefont {Kanwar}}, \bibinfo {author} {\bibfnamefont
  {F.}~\bibnamefont {Romero-L{\'o}pez}}, \bibinfo {author} {\bibfnamefont
  {P.~E.}\ \bibnamefont {Shanahan}},\ and\ \bibinfo {author} {\bibfnamefont
  {J.~M.}\ \bibnamefont {Urban}},\ }\bibfield  {title} {\bibinfo {title}
  {{Variance reduction in lattice QCD observables via normalizing flows}},\
  }\href@noop {} {\bibfield  {journal} {\bibinfo  {journal} {arXiv}\ }
  (\bibinfo {year} {2026})},\ \Eprint {https://arxiv.org/abs/2603.02984}
  {arXiv:2603.02984 [hep-lat]} \BibitemShut {NoStop}%
\bibitem [{\citenamefont {Lucini}\ \emph {et~al.}(2016)\citenamefont {Lucini},
  \citenamefont {Patella}, \citenamefont {Ramos},\ and\ \citenamefont
  {Tantalo}}]{Lucini:2015hfa}%
  \BibitemOpen
  \bibfield  {author} {\bibinfo {author} {\bibfnamefont {B.}~\bibnamefont
  {Lucini}}, \bibinfo {author} {\bibfnamefont {A.}~\bibnamefont {Patella}},
  \bibinfo {author} {\bibfnamefont {A.}~\bibnamefont {Ramos}},\ and\ \bibinfo
  {author} {\bibfnamefont {N.}~\bibnamefont {Tantalo}},\ }\bibfield  {title}
  {\bibinfo {title} {{Charged hadrons in local finite-volume QED+QCD with
  C$^{*}$ boundary conditions}},\ }\href
  {https://doi.org/10.1007/JHEP02(2016)076} {\bibfield  {journal} {\bibinfo
  {journal} {JHEP}\ }\textbf {\bibinfo {volume} {02}},\ \bibinfo {pages}
  {076}},\ \Eprint {https://arxiv.org/abs/1509.01636} {arXiv:1509.01636
  [hep-th]} \BibitemShut {NoStop}%
\bibitem [{\citenamefont {Albandea}\ \emph {et~al.}(2026)\citenamefont
  {Albandea}, \citenamefont {Kuberski},\ and\ \citenamefont
  {Panadero}}]{Albandea:2026cqh}%
  \BibitemOpen
  \bibfield  {author} {\bibinfo {author} {\bibfnamefont {D.}~\bibnamefont
  {Albandea}}, \bibinfo {author} {\bibfnamefont {S.}~\bibnamefont {Kuberski}},\
  and\ \bibinfo {author} {\bibfnamefont {F.~P.}\ \bibnamefont {Panadero}},\
  }\bibfield  {title} {\bibinfo {title} {{A first study of strong isospin
  breaking effects in lattice QCD using truncated polynomials}},\ }in\
  \href@noop {} {\emph {\bibinfo {booktitle} {{42th International Symposium on
  Lattice Field Theory}}}}\ (\bibinfo {year} {2026})\ \Eprint
  {https://arxiv.org/abs/2603.24420} {arXiv:2603.24420 [hep-lat]} \BibitemShut
  {NoStop}%
\bibitem [{\citenamefont {Endres}\ \emph {et~al.}(2016)\citenamefont {Endres},
  \citenamefont {Shindler}, \citenamefont {Tiburzi},\ and\ \citenamefont
  {Walker-Loud}}]{Endres:2015gda}%
  \BibitemOpen
  \bibfield  {author} {\bibinfo {author} {\bibfnamefont {M.~G.}\ \bibnamefont
  {Endres}}, \bibinfo {author} {\bibfnamefont {A.}~\bibnamefont {Shindler}},
  \bibinfo {author} {\bibfnamefont {B.~C.}\ \bibnamefont {Tiburzi}},\ and\
  \bibinfo {author} {\bibfnamefont {A.}~\bibnamefont {Walker-Loud}},\
  }\bibfield  {title} {\bibinfo {title} {{Massive photons: an infrared
  regularization scheme for lattice QCD+QED}},\ }\href
  {https://doi.org/10.1103/PhysRevLett.117.072002} {\bibfield  {journal}
  {\bibinfo  {journal} {Phys. Rev. Lett.}\ }\textbf {\bibinfo {volume} {117}},\
  \bibinfo {pages} {072002} (\bibinfo {year} {2016})},\ \Eprint
  {https://arxiv.org/abs/1507.08916} {arXiv:1507.08916 [hep-lat]} \BibitemShut
  {NoStop}%
\bibitem [{\citenamefont {Boyle}\ \emph {et~al.}(2017)\citenamefont {Boyle},
  \citenamefont {G{\"u}lpers}, \citenamefont {Harrison}, \citenamefont
  {J{\"u}ttner}, \citenamefont {Lehner}, \citenamefont {Portelli},\ and\
  \citenamefont {Sachrajda}}]{Boyle:2017gzv}%
  \BibitemOpen
  \bibfield  {author} {\bibinfo {author} {\bibfnamefont {P.}~\bibnamefont
  {Boyle}}, \bibinfo {author} {\bibfnamefont {V.}~\bibnamefont {G{\"u}lpers}},
  \bibinfo {author} {\bibfnamefont {J.}~\bibnamefont {Harrison}}, \bibinfo
  {author} {\bibfnamefont {A.}~\bibnamefont {J{\"u}ttner}}, \bibinfo {author}
  {\bibfnamefont {C.}~\bibnamefont {Lehner}}, \bibinfo {author} {\bibfnamefont
  {A.}~\bibnamefont {Portelli}},\ and\ \bibinfo {author} {\bibfnamefont
  {C.~T.}\ \bibnamefont {Sachrajda}},\ }\bibfield  {title} {\bibinfo {title}
  {{Isospin breaking corrections to meson masses and the hadronic vacuum
  polarization: a comparative study}},\ }\href
  {https://doi.org/10.1007/JHEP09(2017)153} {\bibfield  {journal} {\bibinfo
  {journal} {JHEP}\ }\textbf {\bibinfo {volume} {09}},\ \bibinfo {pages}
  {153}},\ \Eprint {https://arxiv.org/abs/1706.05293} {arXiv:1706.05293
  [hep-lat]} \BibitemShut {NoStop}%
\bibitem [{\citenamefont {Altherr}\ \emph
  {et~al.}(2026{\natexlab{b}})\citenamefont {Altherr} \emph
  {et~al.}}]{Altherr:2026pnt}%
  \BibitemOpen
  \bibfield  {author} {\bibinfo {author} {\bibfnamefont {A.}~\bibnamefont
  {Altherr}} \emph {et~al.},\ }\bibfield  {title} {\bibinfo {title} {{Comparing
  RM123 and non-perturbative QCD+QED approaches to the HVP with C-periodic
  boundary conditions}},\ }in\ \href@noop {} {\emph {\bibinfo {booktitle}
  {{42th International Symposium on Lattice Field Theory}}}}\ (\bibinfo {year}
  {2026})\ \Eprint {https://arxiv.org/abs/2605.01085} {arXiv:2605.01085
  [hep-lat]} \BibitemShut {NoStop}%
\bibitem [{\citenamefont {Davoudi}\ \emph {et~al.}(2019)\citenamefont
  {Davoudi}, \citenamefont {Harrison}, \citenamefont {J\"uttner}, \citenamefont
  {Portelli},\ and\ \citenamefont {Savage}}]{Davoudi:2018qpl}%
  \BibitemOpen
  \bibfield  {author} {\bibinfo {author} {\bibfnamefont {Z.}~\bibnamefont
  {Davoudi}}, \bibinfo {author} {\bibfnamefont {J.}~\bibnamefont {Harrison}},
  \bibinfo {author} {\bibfnamefont {A.}~\bibnamefont {J\"uttner}}, \bibinfo
  {author} {\bibfnamefont {A.}~\bibnamefont {Portelli}},\ and\ \bibinfo
  {author} {\bibfnamefont {M.~J.}\ \bibnamefont {Savage}},\ }\bibfield  {title}
  {\bibinfo {title} {{Theoretical aspects of quantum electrodynamics in a
  finite volume with periodic boundary conditions}},\ }\href
  {https://doi.org/10.1103/PhysRevD.99.034510} {\bibfield  {journal} {\bibinfo
  {journal} {Phys. Rev. D}\ }\textbf {\bibinfo {volume} {99}},\ \bibinfo
  {pages} {034510} (\bibinfo {year} {2019})},\ \Eprint
  {https://arxiv.org/abs/1810.05923} {arXiv:1810.05923 [hep-lat]} \BibitemShut
  {NoStop}%
\bibitem [{\citenamefont {Bijnens}\ \emph {et~al.}(2019)\citenamefont
  {Bijnens}, \citenamefont {Harrison}, \citenamefont {Hermansson-Truedsson},
  \citenamefont {Janowski}, \citenamefont {J{\"u}ttner},\ and\ \citenamefont
  {Portelli}}]{Bijnens:2019ejw}%
  \BibitemOpen
  \bibfield  {author} {\bibinfo {author} {\bibfnamefont {J.}~\bibnamefont
  {Bijnens}}, \bibinfo {author} {\bibfnamefont {J.}~\bibnamefont {Harrison}},
  \bibinfo {author} {\bibfnamefont {N.}~\bibnamefont {Hermansson-Truedsson}},
  \bibinfo {author} {\bibfnamefont {T.}~\bibnamefont {Janowski}}, \bibinfo
  {author} {\bibfnamefont {A.}~\bibnamefont {J{\"u}ttner}},\ and\ \bibinfo
  {author} {\bibfnamefont {A.}~\bibnamefont {Portelli}},\ }\bibfield  {title}
  {\bibinfo {title} {{Electromagnetic finite-size effects to the hadronic
  vacuum polarization}},\ }\href {https://doi.org/10.1103/PhysRevD.100.014508}
  {\bibfield  {journal} {\bibinfo  {journal} {Phys. Rev. D}\ }\textbf {\bibinfo
  {volume} {100}},\ \bibinfo {pages} {014508} (\bibinfo {year} {2019})},\
  \Eprint {https://arxiv.org/abs/1903.10591} {arXiv:1903.10591 [hep-lat]}
  \BibitemShut {NoStop}%
\bibitem [{\citenamefont {Kronfeld}\ and\ \citenamefont
  {Wiese}(1991)}]{Kronfeld:1990qu}%
  \BibitemOpen
  \bibfield  {author} {\bibinfo {author} {\bibfnamefont {A.~S.}\ \bibnamefont
  {Kronfeld}}\ and\ \bibinfo {author} {\bibfnamefont {U.~J.}\ \bibnamefont
  {Wiese}},\ }\bibfield  {title} {\bibinfo {title} {{SU(N) gauge theories with
  C periodic boundary conditions. 1. Topological structure}},\ }\href
  {https://doi.org/10.1016/0550-3213(91)90479-H} {\bibfield  {journal}
  {\bibinfo  {journal} {Nucl. Phys. B}\ }\textbf {\bibinfo {volume} {357}},\
  \bibinfo {pages} {521} (\bibinfo {year} {1991})}\BibitemShut {NoStop}%
\bibitem [{\citenamefont {Wiese}(1992)}]{Wiese:1991ku}%
  \BibitemOpen
  \bibfield  {author} {\bibinfo {author} {\bibfnamefont {U.~J.}\ \bibnamefont
  {Wiese}},\ }\bibfield  {title} {\bibinfo {title} {{C periodic and G periodic
  QCD at finite temperature}},\ }\href
  {https://doi.org/10.1016/0550-3213(92)90333-7} {\bibfield  {journal}
  {\bibinfo  {journal} {Nucl. Phys. B}\ }\textbf {\bibinfo {volume} {375}},\
  \bibinfo {pages} {45} (\bibinfo {year} {1992})}\BibitemShut {NoStop}%
\bibitem [{\citenamefont {Kronfeld}\ and\ \citenamefont
  {Wiese}(1993)}]{Kronfeld:1992ae}%
  \BibitemOpen
  \bibfield  {author} {\bibinfo {author} {\bibfnamefont {A.~S.}\ \bibnamefont
  {Kronfeld}}\ and\ \bibinfo {author} {\bibfnamefont {U.~J.}\ \bibnamefont
  {Wiese}},\ }\bibfield  {title} {\bibinfo {title} {{SU(N) gauge theories with
  C periodic boundary conditions. 2. Small volume dynamics}},\ }\href
  {https://doi.org/10.1016/0550-3213(93)90302-6} {\bibfield  {journal}
  {\bibinfo  {journal} {Nucl. Phys. B}\ }\textbf {\bibinfo {volume} {401}},\
  \bibinfo {pages} {190} (\bibinfo {year} {1993})},\ \Eprint
  {https://arxiv.org/abs/hep-lat/9210008} {arXiv:hep-lat/9210008} \BibitemShut
  {NoStop}%
\bibitem [{\citenamefont {Polley}(1993)}]{Polley:1993bn}%
  \BibitemOpen
  \bibfield  {author} {\bibinfo {author} {\bibfnamefont {L.}~\bibnamefont
  {Polley}},\ }\bibfield  {title} {\bibinfo {title} {{Boundaries for $SU(3)_C
  \times U(1)_{\textrm{el}}$ lattice gauge theory with a chemical potential}},\
  }\href {https://doi.org/10.1007/BF01555844} {\bibfield  {journal} {\bibinfo
  {journal} {Z. Phys. C}\ }\textbf {\bibinfo {volume} {59}},\ \bibinfo {pages}
  {105} (\bibinfo {year} {1993})}\BibitemShut {NoStop}%
\bibitem [{\citenamefont {Duncan}\ \emph {et~al.}(1997)\citenamefont {Duncan},
  \citenamefont {Eichten},\ and\ \citenamefont {Thacker}}]{Duncan:1996be}%
  \BibitemOpen
  \bibfield  {author} {\bibinfo {author} {\bibfnamefont {A.}~\bibnamefont
  {Duncan}}, \bibinfo {author} {\bibfnamefont {E.}~\bibnamefont {Eichten}},\
  and\ \bibinfo {author} {\bibfnamefont {H.}~\bibnamefont {Thacker}},\
  }\bibfield  {title} {\bibinfo {title} {{Electromagnetic structure of light
  baryons in lattice QCD}},\ }\href
  {https://doi.org/10.1016/S0370-2693(97)00850-2} {\bibfield  {journal}
  {\bibinfo  {journal} {Phys. Lett. B}\ }\textbf {\bibinfo {volume} {409}},\
  \bibinfo {pages} {387} (\bibinfo {year} {1997})},\ \Eprint
  {https://arxiv.org/abs/hep-lat/9607032} {arXiv:hep-lat/9607032} \BibitemShut
  {NoStop}%
\bibitem [{\citenamefont {Duncan}\ \emph {et~al.}(1996)\citenamefont {Duncan},
  \citenamefont {Eichten},\ and\ \citenamefont {Thacker}}]{Duncan:1996xy}%
  \BibitemOpen
  \bibfield  {author} {\bibinfo {author} {\bibfnamefont {A.}~\bibnamefont
  {Duncan}}, \bibinfo {author} {\bibfnamefont {E.}~\bibnamefont {Eichten}},\
  and\ \bibinfo {author} {\bibfnamefont {H.}~\bibnamefont {Thacker}},\
  }\bibfield  {title} {\bibinfo {title} {{Electromagnetic splittings and light
  quark masses in lattice QCD}},\ }\href
  {https://doi.org/10.1103/PhysRevLett.76.3894} {\bibfield  {journal} {\bibinfo
   {journal} {Phys. Rev. Lett.}\ }\textbf {\bibinfo {volume} {76}},\ \bibinfo
  {pages} {3894} (\bibinfo {year} {1996})},\ \Eprint
  {https://arxiv.org/abs/hep-lat/9602005} {arXiv:hep-lat/9602005} \BibitemShut
  {NoStop}%
\bibitem [{\citenamefont {Hayakawa}\ and\ \citenamefont
  {Uno}(2008)}]{Hayakawa:2008an}%
  \BibitemOpen
  \bibfield  {author} {\bibinfo {author} {\bibfnamefont {M.}~\bibnamefont
  {Hayakawa}}\ and\ \bibinfo {author} {\bibfnamefont {S.}~\bibnamefont {Uno}},\
  }\bibfield  {title} {\bibinfo {title} {{QED in finite volume and finite size
  scaling effect on electromagnetic properties of hadrons}},\ }\href
  {https://doi.org/10.1143/PTP.120.413} {\bibfield  {journal} {\bibinfo
  {journal} {Prog. Theor. Phys.}\ }\textbf {\bibinfo {volume} {120}},\ \bibinfo
  {pages} {413} (\bibinfo {year} {2008})},\ \Eprint
  {https://arxiv.org/abs/0804.2044} {arXiv:0804.2044 [hep-ph]} \BibitemShut
  {NoStop}%
\bibitem [{\citenamefont {Asmussen}\ \emph {et~al.}(2016)\citenamefont
  {Asmussen}, \citenamefont {Green}, \citenamefont {Meyer},\ and\ \citenamefont
  {Nyffeler}}]{Asmussen:2016lse}%
  \BibitemOpen
  \bibfield  {author} {\bibinfo {author} {\bibfnamefont {N.}~\bibnamefont
  {Asmussen}}, \bibinfo {author} {\bibfnamefont {J.}~\bibnamefont {Green}},
  \bibinfo {author} {\bibfnamefont {H.~B.}\ \bibnamefont {Meyer}},\ and\
  \bibinfo {author} {\bibfnamefont {A.}~\bibnamefont {Nyffeler}},\ }\bibfield
  {title} {\bibinfo {title} {{Position-space approach to hadronic
  light-by-light scattering in the muon $g-2$ on the lattice}},\ }\href
  {https://doi.org/10.22323/1.256.0164} {\bibfield  {journal} {\bibinfo
  {journal} {PoS}\ }\textbf {\bibinfo {volume} {LATTICE2016}},\ \bibinfo
  {pages} {164} (\bibinfo {year} {2016})},\ \Eprint
  {https://arxiv.org/abs/1609.08454} {arXiv:1609.08454 [hep-lat]} \BibitemShut
  {NoStop}%
\bibitem [{\citenamefont {Feng}\ and\ \citenamefont
  {Jin}(2019)}]{Feng:2018qpx}%
  \BibitemOpen
  \bibfield  {author} {\bibinfo {author} {\bibfnamefont {X.}~\bibnamefont
  {Feng}}\ and\ \bibinfo {author} {\bibfnamefont {L.}~\bibnamefont {Jin}},\
  }\bibfield  {title} {\bibinfo {title} {{QED self energies from lattice QCD
  without power-law finite-volume errors}},\ }\href
  {https://doi.org/10.1103/PhysRevD.100.094509} {\bibfield  {journal} {\bibinfo
   {journal} {Phys. Rev. D}\ }\textbf {\bibinfo {volume} {100}},\ \bibinfo
  {pages} {094509} (\bibinfo {year} {2019})},\ \Eprint
  {https://arxiv.org/abs/1812.09817} {arXiv:1812.09817 [hep-lat]} \BibitemShut
  {NoStop}%
\bibitem [{\citenamefont {Di~Carlo}\ \emph {et~al.}(2025)\citenamefont
  {Di~Carlo}, \citenamefont {Hansen}, \citenamefont {Hermansson-Truedsson},\
  and\ \citenamefont {Portelli}}]{DiCarlo:2025uyj}%
  \BibitemOpen
  \bibfield  {author} {\bibinfo {author} {\bibfnamefont {M.}~\bibnamefont
  {Di~Carlo}}, \bibinfo {author} {\bibfnamefont {M.~T.}\ \bibnamefont
  {Hansen}}, \bibinfo {author} {\bibfnamefont {N.}~\bibnamefont
  {Hermansson-Truedsson}},\ and\ \bibinfo {author} {\bibfnamefont
  {A.}~\bibnamefont {Portelli}},\ }\bibfield  {title} {\bibinfo {title}
  {{QED$_{r}$: a finite-volume QED action with redistributed spatial
  zero-momentum modes}},\ }\href {https://doi.org/10.1007/JHEP10(2025)026}
  {\bibfield  {journal} {\bibinfo  {journal} {JHEP}\ }\textbf {\bibinfo
  {volume} {10}},\ \bibinfo {pages} {026}},\ \Eprint
  {https://arxiv.org/abs/2501.07936} {arXiv:2501.07936 [hep-lat]} \BibitemShut
  {NoStop}%
\bibitem [{\citenamefont {Cranmer}\ \emph {et~al.}(2023)\citenamefont
  {Cranmer}, \citenamefont {Kanwar}, \citenamefont {Racani{\`e}re},
  \citenamefont {Rezende},\ and\ \citenamefont {Shanahan}}]{Cranmer:2023xbe}%
  \BibitemOpen
  \bibfield  {author} {\bibinfo {author} {\bibfnamefont {K.}~\bibnamefont
  {Cranmer}}, \bibinfo {author} {\bibfnamefont {G.}~\bibnamefont {Kanwar}},
  \bibinfo {author} {\bibfnamefont {S.}~\bibnamefont {Racani{\`e}re}}, \bibinfo
  {author} {\bibfnamefont {D.~J.}\ \bibnamefont {Rezende}},\ and\ \bibinfo
  {author} {\bibfnamefont {P.~E.}\ \bibnamefont {Shanahan}},\ }\bibfield
  {title} {\bibinfo {title} {{Advances in machine-learning-based sampling
  motivated by lattice quantum chromodynamics}},\ }\href
  {https://doi.org/10.1038/s42254-023-00616-w} {\bibfield  {journal} {\bibinfo
  {journal} {Nature Rev. Phys.}\ }\textbf {\bibinfo {volume} {5}},\ \bibinfo
  {pages} {526} (\bibinfo {year} {2023})},\ \Eprint
  {https://arxiv.org/abs/2309.01156} {arXiv:2309.01156 [hep-lat]} \BibitemShut
  {NoStop}%
\bibitem [{\citenamefont {Kanwar}(2024)}]{Kanwar:2024ujc}%
  \BibitemOpen
  \bibfield  {author} {\bibinfo {author} {\bibfnamefont {G.}~\bibnamefont
  {Kanwar}},\ }\bibfield  {title} {\bibinfo {title} {{Flow-based sampling for
  lattice field theories}},\ }in\ \href@noop {} {\emph {\bibinfo {booktitle}
  {{40th International Symposium on Lattice Field Theory}}}}\ (\bibinfo {year}
  {2024})\ \Eprint {https://arxiv.org/abs/2401.01297} {arXiv:2401.01297
  [hep-lat]} \BibitemShut {NoStop}%
\bibitem [{\citenamefont {Wenger}(2026)}]{Wenger:2026sjp}%
  \BibitemOpen
  \bibfield  {author} {\bibinfo {author} {\bibfnamefont {U.}~\bibnamefont
  {Wenger}},\ }\bibfield  {title} {\bibinfo {title} {{Machine learning for
  four-dimensional SU(3) lattice gauge theories}},\ }in\ \href@noop {} {\emph
  {\bibinfo {booktitle} {{42th International Symposium on Lattice Field
  Theory}}}}\ (\bibinfo {year} {2026})\ \Eprint
  {https://arxiv.org/abs/2604.12416} {arXiv:2604.12416 [hep-lat]} \BibitemShut
  {NoStop}%
\bibitem [{\citenamefont {Chen}\ \emph {et~al.}(2019)\citenamefont {Chen},
  \citenamefont {Behrmann}, \citenamefont {Duvenaud},\ and\ \citenamefont
  {Jacobsen}}]{chen2019residual}%
  \BibitemOpen
  \bibfield  {author} {\bibinfo {author} {\bibfnamefont {R.~T.}\ \bibnamefont
  {Chen}}, \bibinfo {author} {\bibfnamefont {J.}~\bibnamefont {Behrmann}},
  \bibinfo {author} {\bibfnamefont {D.~K.}\ \bibnamefont {Duvenaud}},\ and\
  \bibinfo {author} {\bibfnamefont {J.-H.}\ \bibnamefont {Jacobsen}},\
  }\bibfield  {title} {\bibinfo {title} {Residual flows for invertible
  generative modeling},\ }\href@noop {} {\bibfield  {journal} {\bibinfo
  {journal} {Advances in neural information processing systems}\ }\textbf
  {\bibinfo {volume} {32}} (\bibinfo {year} {2019})},\ \Eprint
  {https://arxiv.org/abs/1906.02735} {arXiv:1906.02735 [stat.ML]} \BibitemShut
  {NoStop}%
\bibitem [{\citenamefont {Ronneberger}\ \emph {et~al.}(2015)\citenamefont
  {Ronneberger}, \citenamefont {Fischer},\ and\ \citenamefont
  {Brox}}]{ronneberger2015u}%
  \BibitemOpen
  \bibfield  {author} {\bibinfo {author} {\bibfnamefont {O.}~\bibnamefont
  {Ronneberger}}, \bibinfo {author} {\bibfnamefont {P.}~\bibnamefont
  {Fischer}},\ and\ \bibinfo {author} {\bibfnamefont {T.}~\bibnamefont
  {Brox}},\ }\bibfield  {title} {\bibinfo {title} {U-net: Convolutional
  networks for biomedical image segmentation},\ }in\ \href@noop {} {\emph
  {\bibinfo {booktitle} {International Conference on Medical image computing
  and computer-assisted intervention}}}\ (\bibinfo {organization} {Springer},\
  \bibinfo {year} {2015})\ pp.\ \bibinfo {pages} {234--241},\ \Eprint
  {https://arxiv.org/abs/1505.04597} {arXiv:1505.04597 [cs.CV]} \BibitemShut
  {NoStop}%
\bibitem [{\citenamefont {Kullback}\ and\ \citenamefont
  {Leibler}(1951)}]{Kullback:1951zyt}%
  \BibitemOpen
  \bibfield  {author} {\bibinfo {author} {\bibfnamefont {S.}~\bibnamefont
  {Kullback}}\ and\ \bibinfo {author} {\bibfnamefont {R.~A.}\ \bibnamefont
  {Leibler}},\ }\bibfield  {title} {\bibinfo {title} {{On Information and
  Sufficiency}},\ }\href {https://doi.org/10.1214/aoms/1177729694} {\bibfield
  {journal} {\bibinfo  {journal} {The Annals of Mathematical Statistics}\
  }\textbf {\bibinfo {volume} {22}},\ \bibinfo {pages} {79} (\bibinfo {year}
  {1951})}\BibitemShut {NoStop}%
\bibitem [{\citenamefont {Vaitl}\ \emph {et~al.}(2022)\citenamefont {Vaitl},
  \citenamefont {Nicoli}, \citenamefont {Nakajima},\ and\ \citenamefont
  {Kessel}}]{vaitl2022gradients}%
  \BibitemOpen
  \bibfield  {author} {\bibinfo {author} {\bibfnamefont {L.}~\bibnamefont
  {Vaitl}}, \bibinfo {author} {\bibfnamefont {K.~A.}\ \bibnamefont {Nicoli}},
  \bibinfo {author} {\bibfnamefont {S.}~\bibnamefont {Nakajima}},\ and\
  \bibinfo {author} {\bibfnamefont {P.}~\bibnamefont {Kessel}},\ }\href@noop {}
  {\bibinfo {title} {Gradients should stay on path: Better estimators of the
  reverse- and forward kl divergence for normalizing flows}} (\bibinfo {year}
  {2022}),\ \Eprint {https://arxiv.org/abs/2207.08219} {arXiv:2207.08219
  [cs.LG]} \BibitemShut {NoStop}%
\bibitem [{\citenamefont {Boyda}\ \emph {et~al.}(2021)\citenamefont {Boyda},
  \citenamefont {Kanwar}, \citenamefont {Racani{\`e}re}, \citenamefont
  {Rezende}, \citenamefont {Albergo}, \citenamefont {Cranmer}, \citenamefont
  {Hackett},\ and\ \citenamefont {Shanahan}}]{Boyda:2020hsi}%
  \BibitemOpen
  \bibfield  {author} {\bibinfo {author} {\bibfnamefont {D.}~\bibnamefont
  {Boyda}}, \bibinfo {author} {\bibfnamefont {G.}~\bibnamefont {Kanwar}},
  \bibinfo {author} {\bibfnamefont {S.}~\bibnamefont {Racani{\`e}re}}, \bibinfo
  {author} {\bibfnamefont {D.~J.}\ \bibnamefont {Rezende}}, \bibinfo {author}
  {\bibfnamefont {M.~S.}\ \bibnamefont {Albergo}}, \bibinfo {author}
  {\bibfnamefont {K.}~\bibnamefont {Cranmer}}, \bibinfo {author} {\bibfnamefont
  {D.~C.}\ \bibnamefont {Hackett}},\ and\ \bibinfo {author} {\bibfnamefont
  {P.~E.}\ \bibnamefont {Shanahan}},\ }\bibfield  {title} {\bibinfo {title}
  {{Sampling using $SU(N)$ gauge equivariant flows}},\ }\href
  {https://doi.org/10.1103/PhysRevD.103.074504} {\bibfield  {journal} {\bibinfo
   {journal} {Phys. Rev. D}\ }\textbf {\bibinfo {volume} {103}},\ \bibinfo
  {pages} {074504} (\bibinfo {year} {2021})},\ \Eprint
  {https://arxiv.org/abs/2008.05456} {arXiv:2008.05456 [hep-lat]} \BibitemShut
  {NoStop}%
\bibitem [{\citenamefont {Albergo}\ \emph {et~al.}(2022)\citenamefont
  {Albergo}, \citenamefont {Boyda}, \citenamefont {Cranmer}, \citenamefont
  {Hackett}, \citenamefont {Kanwar}, \citenamefont {Racani{\`e}re},
  \citenamefont {Rezende}, \citenamefont {Romero-L{\'o}pez}, \citenamefont
  {Shanahan},\ and\ \citenamefont {Urban}}]{Albergo:2022qfi}%
  \BibitemOpen
  \bibfield  {author} {\bibinfo {author} {\bibfnamefont {M.~S.}\ \bibnamefont
  {Albergo}}, \bibinfo {author} {\bibfnamefont {D.}~\bibnamefont {Boyda}},
  \bibinfo {author} {\bibfnamefont {K.}~\bibnamefont {Cranmer}}, \bibinfo
  {author} {\bibfnamefont {D.~C.}\ \bibnamefont {Hackett}}, \bibinfo {author}
  {\bibfnamefont {G.}~\bibnamefont {Kanwar}}, \bibinfo {author} {\bibfnamefont
  {S.}~\bibnamefont {Racani{\`e}re}}, \bibinfo {author} {\bibfnamefont {D.~J.}\
  \bibnamefont {Rezende}}, \bibinfo {author} {\bibfnamefont {F.}~\bibnamefont
  {Romero-L{\'o}pez}}, \bibinfo {author} {\bibfnamefont {P.~E.}\ \bibnamefont
  {Shanahan}},\ and\ \bibinfo {author} {\bibfnamefont {J.~M.}\ \bibnamefont
  {Urban}},\ }\bibfield  {title} {\bibinfo {title} {{Flow-based sampling in the
  lattice Schwinger model at criticality}},\ }\href
  {https://doi.org/10.1103/PhysRevD.106.014514} {\bibfield  {journal} {\bibinfo
   {journal} {Phys. Rev. D}\ }\textbf {\bibinfo {volume} {106}},\ \bibinfo
  {pages} {014514} (\bibinfo {year} {2022})},\ \Eprint
  {https://arxiv.org/abs/2202.11712} {arXiv:2202.11712 [hep-lat]} \BibitemShut
  {NoStop}%
\bibitem [{\citenamefont {Abbott}\ \emph {et~al.}(2022)\citenamefont {Abbott}
  \emph {et~al.}}]{Abbott:2022zhs}%
  \BibitemOpen
  \bibfield  {author} {\bibinfo {author} {\bibfnamefont {R.}~\bibnamefont
  {Abbott}} \emph {et~al.},\ }\bibfield  {title} {\bibinfo {title}
  {{Gauge-equivariant flow models for sampling in lattice field theories with
  pseudofermions}},\ }\href {https://doi.org/10.1103/PhysRevD.106.074506}
  {\bibfield  {journal} {\bibinfo  {journal} {Phys. Rev. D}\ }\textbf {\bibinfo
  {volume} {106}},\ \bibinfo {pages} {074506} (\bibinfo {year} {2022})},\
  \Eprint {https://arxiv.org/abs/2207.08945} {arXiv:2207.08945 [hep-lat]}
  \BibitemShut {NoStop}%
\bibitem [{\citenamefont {McIntosh-Smith}\ \emph {et~al.}(2024)\citenamefont
  {McIntosh-Smith}, \citenamefont {Alam},\ and\ \citenamefont
  {Woods}}]{mcintoshsmith2024isambardaileadershipclasssupercomputer}%
  \BibitemOpen
  \bibfield  {author} {\bibinfo {author} {\bibfnamefont {S.}~\bibnamefont
  {McIntosh-Smith}}, \bibinfo {author} {\bibfnamefont {S.~R.}\ \bibnamefont
  {Alam}},\ and\ \bibinfo {author} {\bibfnamefont {C.}~\bibnamefont {Woods}},\
  }\href@noop {} {\bibinfo {title} {Isambard-{AI}: a leadership class
  supercomputer optimised specifically for artificial intelligence}} (\bibinfo
  {year} {2024}),\ \Eprint {https://arxiv.org/abs/2410.11199} {arXiv:2410.11199
  [cs.DC]} \BibitemShut {NoStop}%
\bibitem [{\citenamefont {Skilling}(1989)}]{skilling1989eigenvalues}%
  \BibitemOpen
  \bibfield  {author} {\bibinfo {author} {\bibfnamefont {J.}~\bibnamefont
  {Skilling}},\ }\bibfield  {title} {\bibinfo {title} {The eigenvalues of
  mega-dimensional matrices},\ }in\ \href
  {https://doi.org/10.1007/978-94-015-7860-8_48} {\emph {\bibinfo {booktitle}
  {Maximum Entropy and Bayesian Methods: Cambridge, England, 1988}}}\ (\bibinfo
   {publisher} {Springer},\ \bibinfo {year} {1989})\ pp.\ \bibinfo {pages}
  {455--466}\BibitemShut {NoStop}%
\bibitem [{\citenamefont {Hutchinson}(1989)}]{hutchinson1989stochastic}%
  \BibitemOpen
  \bibfield  {author} {\bibinfo {author} {\bibfnamefont {M.~F.}\ \bibnamefont
  {Hutchinson}},\ }\bibfield  {title} {\bibinfo {title} {A stochastic estimator
  of the trace of the influence matrix for laplacian smoothing splines},\
  }\href {https://doi.org/10.1080/03610918908812806} {\bibfield  {journal}
  {\bibinfo  {journal} {Communications in Statistics-Simulation and
  Computation}\ }\textbf {\bibinfo {volume} {18}},\ \bibinfo {pages} {1059}
  (\bibinfo {year} {1989})}\BibitemShut {NoStop}%
\end{thebibliography}%

\clearpage
\newpage
\setcounter{equation}{0}
\setcounter{figure}{0}
\setcounter{table}{0}
\setcounter{page}{1}
\makeatletter
\renewcommand{\theequation}{S\arabic{equation}}
\renewcommand{\thefigure}{S\arabic{figure}}
\onecolumngrid
\begin{center}
\textbf{\Large Supplemental Material}
\end{center}

\section*{\label{sec:lpt} Lattice perturbation theory}
Here we numerically determine the electromagnetic mass shift and corresponding two-point function $\delta _{e^2} C_e(t)$ in lattice perturbation theory at order $e^2$. This was already considered for four-dimensional scalar $\qedl$ in finite spatial volume in Refs.~\cite{Davoudi:2018qpl,Bijnens:2019ejw}. We do the calculation for $N_d=2,3,4$ and also take into account the finite temporal extent of the lattice.

\begin{figure}[h]
\includegraphics[width=.4\textwidth]{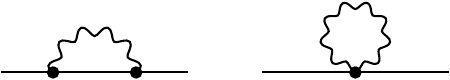}
\caption{The sunset diagram (left) and tadpole diagram (right) contributing to the mass shift at order $e^2$. \label{fig:lptdiagrams}}
\end{figure}

We first expand the operator $\Delta$ in the Lagrangian in \eqref{eq:lagrangian} according to $\Delta = \Delta _0 + e\, \Delta _1 + e^2\, \Delta _2$. It is straightforward to extract the corresponding Feynman rules, and from these one can calculate the two diagrams in Fig.~\ref{fig:lptdiagrams}, here denoted the sunset and tadpole topologies. These diagrams define the central object here, the electromagnetic self energy $\Sigma (p)$, which is given for incoming momentum $p$ by 
\begin{align}\label{eq:selfenergy}
   \Sigma (p) & =  \frac{e^2}{T L^{N_d-1}} \, \sum _{k_0} \sum _{\mathbf{k}\neq \boldsymbol{0}} 
    \Bigg\{
         \frac{N_d-1 + \cos (a p_0)}{\hat{k}^2}
-\frac{\widehat{2p-k} ^2}{\hat{k}^2 \Big( \widehat{p-k}^2+m_0^2\Big) }
    \Bigg\} \, . 
\end{align}
Here the tadpole diagram corresponds to the first term in the summand, and the sunset the second. The hatted momenta are of the standard form $\hat{q}_\mu = (2/a)\, \sin (aq_\mu /2 )$. The mass shift is given by $\delta _{e^2}m^2 = - \Sigma (p_{\textrm{os}})$ where the on-shell momentum satisfies $\hat{p}^2_{\textrm{os}} = -m^2_0$. We choose the rest frame of the scalar where $p= (p_0 , \boldsymbol{0})$. The $\qedl$ regularization here is manifest from the absence of the zero-mode $\mathbf{k}=\boldsymbol{0}$ in the spatial sum. The momenta are quantized according to $ k_0 = (2\pi n_0)/ T$ and $ k_i = (2\pi n_i)/ L$, where the integers $n_0$ and $n_i$ are bounded by $-\hat{T}/2\leq n_0 < \hat{T}/2 $ and $-\hat{L}/2 \leq n_i < \hat{L}/2$. This can be numerically evaluated by either performing the sums explicitly, or by rewriting the sum as an integral as in Ref.~\cite{Bijnens:2019ejw}. 

As laid out in Ref.~\cite{Davoudi:2018qpl}, the order-$e^2$ correlator $\delta _{e^2}C_e(t)$ is defined in terms of the self energy $\Sigma (p) $. With the finite time extent here, the shifted and the tree-level correlator are
\begin{align} \label{eq:lpt-corr}
    \delta _{e^2}C_e(t) 
    & =
    \frac{1}{T} \sum _{p_0} \frac{\Sigma (p_0,\boldsymbol{0})}{(\hat{p}_0^2+m^2)^2}\, e^{i p_0 t} \, ,
    \\
    C_{0} (t) 
    & =
    \frac{1}{T} \sum _{p_0} \frac{1}{\hat{p}_0^2+m^2}\, e^{i p_0 t} \, . 
\end{align}
These are easily numerically evaluated as an inverse discrete Fast Fourier Transform. 

\section*{Flow-model architecture}

Throughout this work, we use a residual flow architecture~\cite{chen2019residual} for the machine-learned normalizing flow transformation.
In terms of the fields $\phi$ and $A$, this architecture is described by the simple additive transformations
\begin{equation} \label{eq:resid-layer}
    \phi' = \phi + \epsilon \, b_\phi(\phi, A)
    \quad \text{and} \quad
    A' = A + \epsilon \, P[b_A(\phi, A)],
\end{equation}
where $P[\cdot] \equiv P_{\qedl}[\cdot]$ indicates projection according to the $\qedl$ constraint. For the purpose of this transformation, we can generally consider $\phi$ and $A$ as general real vectors with $2N$ and $N_d N$ components, respectively, in terms of the total number of sites $N = |\Lambda^{N_d}|$ of the lattice. The additive terms $b_\phi: \mathbb{R}^{(2+N_d)N} \to \mathbb{R}^{2 N}$ and $b_A: \mathbb{R}^{(2+N_d)N} \to \mathbb{R}^{N_d N}$ can then be parameterized by a variety of possible methods. Here we work with a U-net representation~\cite{ronneberger2015u} with one layer of coarsening, with internal transformations using convolutional networks consisting of one hidden layer with $32$ channels.

Under the condition that $\epsilon b_\phi$ and $\epsilon P[b_A]$ are Lipschitz-1, this yields an invertible transformation, although the inverse function has no closed form. The inverse function is not required for training via the standard KL divergence, but is needed when training with path gradients (see below) or for checks of the model. In practice, we also implement the inverse transformation of \eqref{eq:resid-layer} by the fixed-point iteration
\begin{equation}
    \phi_{i+1} = \phi' - \epsilon \, b_\phi (\phi_i, A_i)
    \quad \text{and} \quad
    A_{i+1} = A' - \epsilon \, P[b_A(\phi_i, A_i)].
\end{equation}
Under the same Lipschitz-1 constraint (i.e., given that $\epsilon |b|$ is small in both equations above), this process converges exponentially to the true fixed point solution $(\phi, A)$. Numerically, we find good convergence of the inverse iteration in all runs, indicating that the trained models remain invertible. We note that explicit constructions of layers with Lipschitz-1 guarantees have also been proposed~\cite{chen2019residual}, and it would be interesting to enforce invertibility exactly following such methods in future work.

The Jacobian of the transformation in \eqref{eq:resid-layer} can be obtained from automatic differentiation by brute force evaluation of all components:
\begin{equation} \label{eq:resid-jac}
    J = \begin{pmatrix}
        I + \epsilon \, \nabla_\phi b_\phi & \epsilon \,  \nabla_{P[A]} b_\phi \\
        \epsilon \, P[\nabla_{\phi} b_A] & I + \epsilon \, P[\nabla_{P[A]} b_A]
    \end{pmatrix}.
\end{equation}
Because the $\qedl$ projection is linear, the projected outputs and gradients inside the Jacobian amount to computing
\begin{equation} \label{eq:resid-detJ}
    \det J = \det \left(I + P \epsilon J_b P \right) = \det \left(P (I + \epsilon J_b) P + (I-P) \right)
    = \det_P(I + \epsilon J_b),
\end{equation}
where the final term indicates the determinant taken inside the projected subspace. In practice, it is convenient to evaluate the Jacobian using the unprojected gradient $\nabla_A$ instead of $\nabla_{P[A]}$, as the resulting Jacobian matrix has an identical determinant to the properly projected Jacobian:
\begin{equation}
    \det(I + \epsilon P J_b) = \det(I + \epsilon P J_b P).
\end{equation}
This identity has been checked numerically in the implementation for the numerical results presented in this work.

Evaluating the Jacobian determinant exactly according to \eqref{eq:resid-jac} and \eqref{eq:resid-detJ} is sufficiently tractable for training in small volumes, but quickly becomes prohibitively expensive for the larger physical volumes used during evaluation. As such, when evaluating the machine-learned flow for all lattices larger than or equal to the training geometry, a stochastic Jacobian estimator is used. Following Ref.~\cite{chen2019residual}, a series expansion can be used to replace the determinant with a sum over traces as
\begin{equation}
    \log \det J = \sum_{k=1}^{\infty} \frac{(-1)^{k+1}}{k} \mathrm{tr}(J^k).
\end{equation}
The traces up to a given order can the be evaluated stochastically by an unbiased estimator~\cite{skilling1989eigenvalues,hutchinson1989stochastic}
\begin{equation}
    \log \det J = \sum_{k=1}^{\infty} \frac{(-1)^{k+1}}{k} \mathbb{E}_{\eta}[\eta^\dagger J^k \eta],
\end{equation}
where in each trace the noise vector $\eta$ consists of a number of components equal to the total components of the fields $(\phi, A)$ and is sampled from an isotropic Gaussian distribution. Other noise distributions can be used, but we do not explore them further in this work. For the purposes of evaluation, the series is evaluated up to $k = 5$ using respectively $512$ stochastic evaluations per configuration ($N_d = 2,3$) or $256$ stochastic evaluations per configuration ($N_d = 4$).

\section*{Flow-model training}
We denote the flow transformation defined above by $(\phi', A') = f_\theta(\phi, A)$, where $\theta$ indicates the collection of all free parameters defining the flow model; in this case $\theta$ consists of all convolution kernel weights and biases. The log-density of the model distribution produced by the change of measure under this flow is defined by
\begin{equation}
    \log q(\phi', A') = -S_0(\phi) - S_A(A) - \log \det J_f - \log Z_0
\end{equation}
while the target distribution expanded to $O(e^2)$ is
\begin{equation}
    \log p_{e^2}(\phi, A) = -S_0(\phi) -S_A(A) - e\sintdern{1} - e^2 \sintdern{2} - \log Z_e.
\end{equation}
This truncated version of the distribution has been used as the target for optimizing all models demonstrated in this work, using the training method detailed below; while this does not affect the correctness of the all-orders results, it is possible that optimizing the machine-learned flows directly for the all-orders target could improve performance.
This choice will be investigated in future work.

As the normalizing constants are not known exactly, we choose to optimize the model parameters by minimizing the unnormalized Kullback-Leibler divergence~\cite{Kullback:1951zyt} in this case:
\begin{equation} \label{eq:loss-kl}
\begin{aligned}
    \mathcal{L}_{kl} &= \left< \log q(\phi', A') - \log p_{e^2}(\phi', A') - \log(Z_e/Z_0) \right>_{0} \\
    &= \left< \underbrace{- S_0(\phi) - S_A(A)}_{\equiv -S_{\rm uc}(\phi, A)} - \log J_f(\phi, A) + \underbrace{S_0(\phi') + S_A(A')}_{= S_{\rm uc}(\phi', A')} + e \sintdernp{1} + e^2 \sintdernp{2} \right>_0 \geq -\log(Z_e / Z),
\end{aligned}
\end{equation}
where $\left< \cdot \right>_0$ indicates expectation with respect to the fields $(\phi, A)$ drawn according to the uncoupled action $S_{\rm uc}(\phi, A) = S_0(\phi) + S_A(A)$, while the fields fields $(\phi', A') = f_\theta(\phi, A)$ are produced by acting with the flow model on those fields. All quantities in \eqref{eq:loss-kl} are directly computable on mini-batches of free theory samples by applying the flow and using the definition of the action and the tractable model Jacobian.

It is possible to train the model by directly applying auto-differentiation to the estimator in \eqref{eq:loss-kl} for stochastic gradient descent optimization of the parameters $\theta$. However, this often suffers from significant statistical noise arising from the expectation value in the definition of the loss. Ref.~\cite{vaitl2022gradients} and related works have advocated instead for a reduced-variance \emph{path gradient} estimator for the gradient of \eqref{eq:loss-kl}. To define the path gradients, we first introduce the abbreviated notation $x \equiv (\phi, A)$ to indicate dependence on the combination of both scalar and photon fields. Mathematically, the path gradients approach can then be defined by optimizing instead the loss function
\begin{equation} \label{eq:loss-pg}
    \mathcal{L}_{pg} = \left< - S_{\rm uc}(f^{-1}_{\bsquare}(f_{\theta}(x))) + \log J_{f^{-1}_{\bsquare}}(f_{\theta}(x)) + S_{\rm uc}(f_\theta(x)) + e S^{(1)}_{\rm int}(f_\theta(x)) + e^2 S^{(2)}_{\rm int}(f_\theta(x)) \right>_0.
\end{equation}
With respect to \eqref{eq:loss-kl}, only the first two terms arising from the model density $\log q$ are modified. Here the notation $f^{-1}_{\bsquare}$ indicates the application of the inverse flow treated as a \emph{fixed} transformation, independent of model parameters, such that the only dependence on the model parameters is carried through the value of the output field configuration $(\phi', A') = f_\theta(\phi, A)$. It has been shown that the gradient of \eqref{eq:loss-pg} with respect to $\theta$ only differs from the gradient of \eqref{eq:loss-kl} by a total derivative that vanishes inside the expectation value $\left< \cdot \right>_0$. In practice, $f^{-1}_{\bsquare}$ and $J_{f^{-1}_{\bsquare}}$ can be evaluated by applying the model in reverse with model gradients suppressed. A notable benefit of optimization using path gradients is that statistical noise in the gradients approaches zero as the model distribution converges to the target, leading to particularly efficient training when one is interested in flows that transform between similar distributions, as is the case here. As such, this path gradient optimization is used for all machine-learned flows studied in this work.

\section*{Additional numerical comparisons}
\begin{figure}
    \centering
    \includegraphics{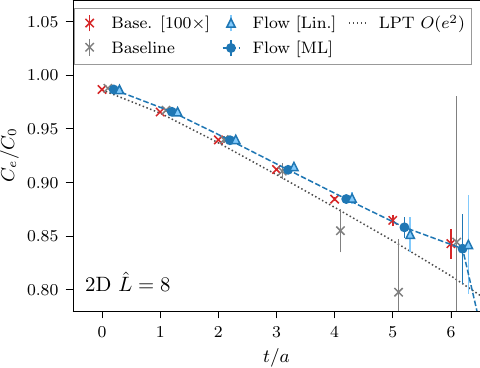} \hspace{0.25cm}
    \includegraphics{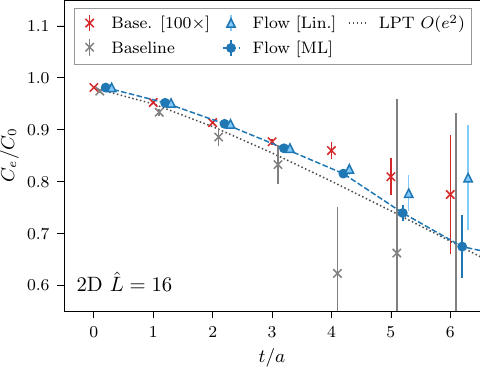}

    \includegraphics{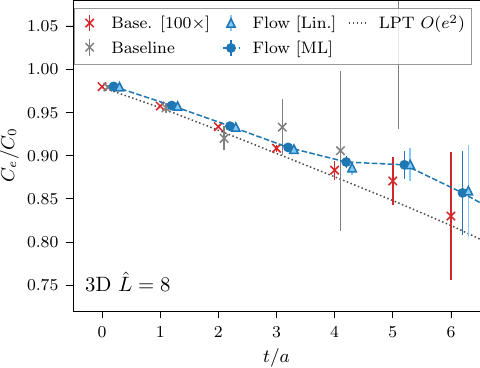} \hspace{0.25cm}
    \includegraphics{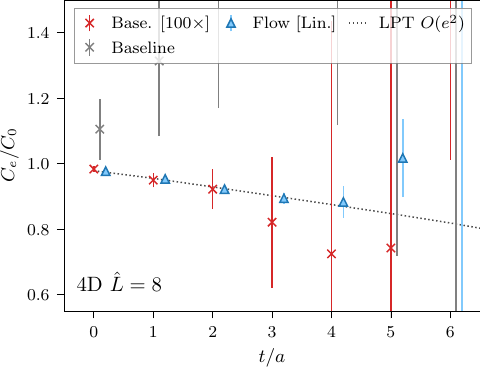}
    \caption{The correlator ratio $C_e(t)/C_0(t)$ in $N_d = 2, 3, 4$ at $\hat{L} = 8, 16$.}
    \label{fig:corr_ratio}
\end{figure}

\begin{figure}[p]
    \centering
    \includegraphics{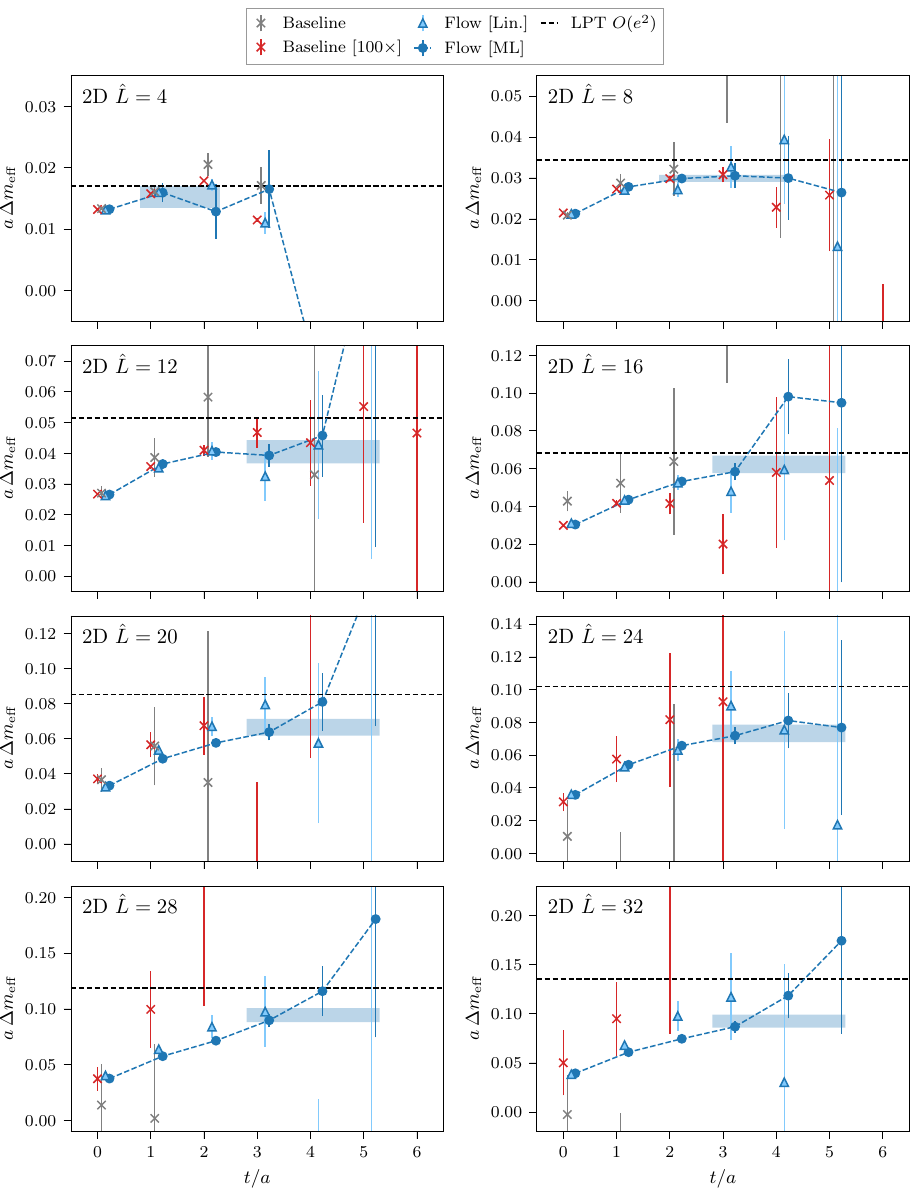}
    \caption{The all-orders effective mass shift in $N_d = 2$ for various $\hat{L}$. \label{fig:effectivemass_varyL_2D}}
\end{figure}

\begin{figure}[p]
    \centering
    \includegraphics{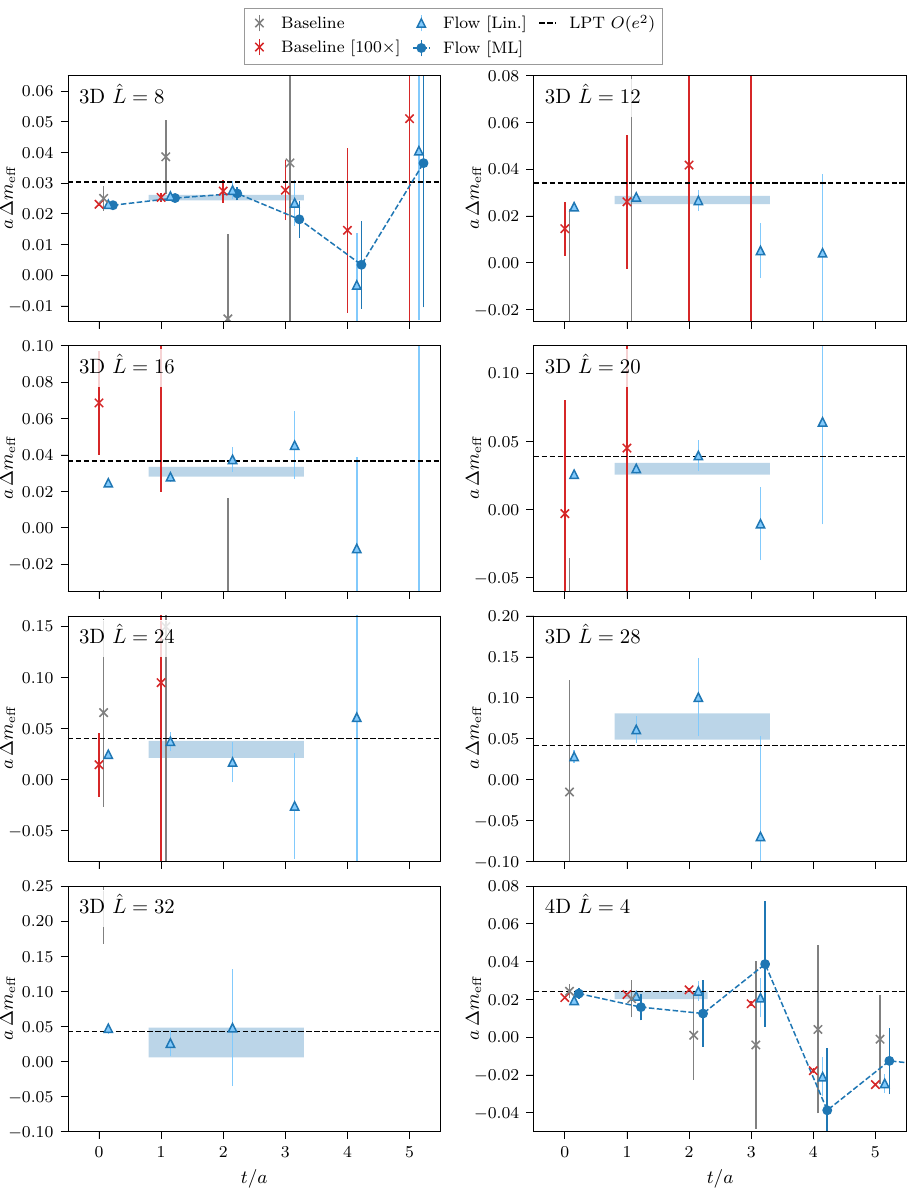}
    \caption{The all-orders effective mass shift in $N_d = 3, 4$ for various $\hat{L}$. \label{fig:effectivemass_varyL_3D}}
\end{figure}

In this section, we provide additional detailed numerical comparisons of the flow method versus existing approaches. The first part of this section highlights further results for observables computed with the QED interaction included to all orders. Fig.~\ref{fig:corr_ratio} shows the correlator ratio $C_{e}(t)/C_{0}(t)$, again for the ML flow against the baseline at the same and higher statistics, as well as the LPT correlator at order $e^2$. It is clear that the flow generally outperforms the baseline at fixed statistics.  
In Figs.~\ref{fig:effectivemass_varyL_2D}--\ref{fig:effectivemass_varyL_3D} the effective mass in $N_d=2,3$ is shown for the flows against the baseline results at both matched statistics and 100 times the statistics. The blue band corresponds to the effective mass obtained from fitting $C_e(t)/C_0(t)$ for the ML flow to an exponential form, as described in the main text. The fit range used is indicated by the time interval, slightly enlarged for visualization. Again, the flowed approach generally performs better than the baseline at fixed statistics. The high-statistics baseline also converges to the flowed results. As we note in the main text, there are small deviations to the order-$e^2$ LPT result.

As opposed to the all-orders calculations considered in the main text, expectation values $\obspe$ can also be calculated at a fixed order in a truncated expansion.
In this case, for both the baseline and flows methods, expectation values are computed with respect to the truncated action
\begin{align}
    \label{eq:sbar}
    \sbar & = 
    S_{0}(\phi) + S_A (A) 
+ e \, \sintdern{1} + e^2\, \sintdern{2}   
    \, ,
\end{align}
which corresponds to $\sfull$ expanded to second order in $e^2$. To evaluate observables at this order using the normalizing flow method, the truncated reweighting factors $\hat{w}_{e^2}(\phi', A')$ are used for all numerical evaluations, where
\begin{equation}
    \hat{w}_{e^2}(\phi', A') \equiv e^{-\bar{S}(\phi', A'; e)}{e^{S_0(\phi) + S_A(A)}} |\det J_f|.
\end{equation}
Because the only modification is the definition of the reweighting factors, these results have been obtained simultaneously with the all-orders results presented in the main text at no additional cost.

\begin{figure}[p]
    \centering
    \includegraphics{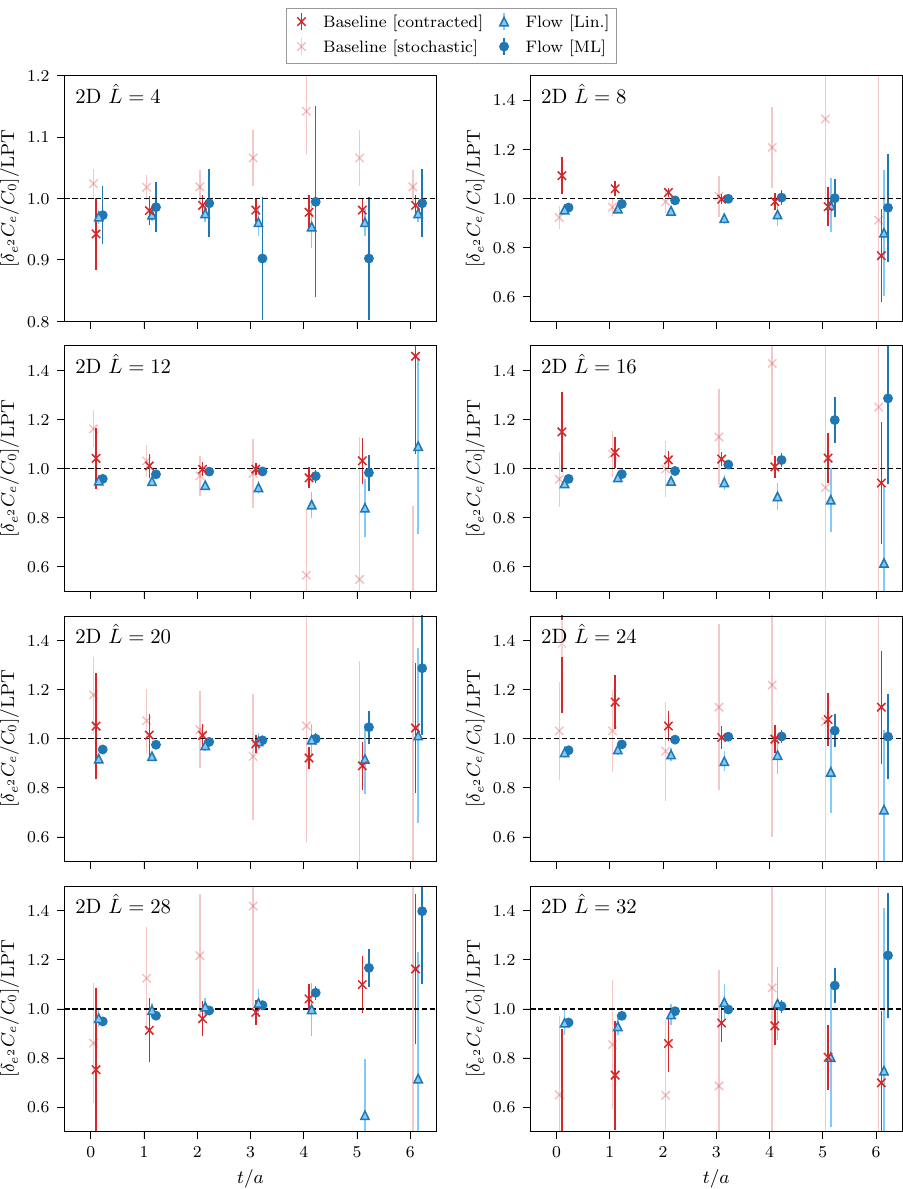}
    \caption{The $O(e^2)$ effective mass shift divided by the analytical LPT result in $N_d = 2$ for various $\hat{L}$. The baseline results are evaluated in the RM123 approach, using either a stochastically sampled photon field (``stochastic'') or the Wick contraction with an exact photon propagator (``contracted''). The flow results are based on a finite-difference estimate of the $O(e^2)$ derivative, and therefore includes an $O(e^4)$ bias as discussed in the text. \label{fig:effectivemass_varyL_2D_e2}}
\end{figure}

\begin{figure}[p]
    \centering
    \includegraphics{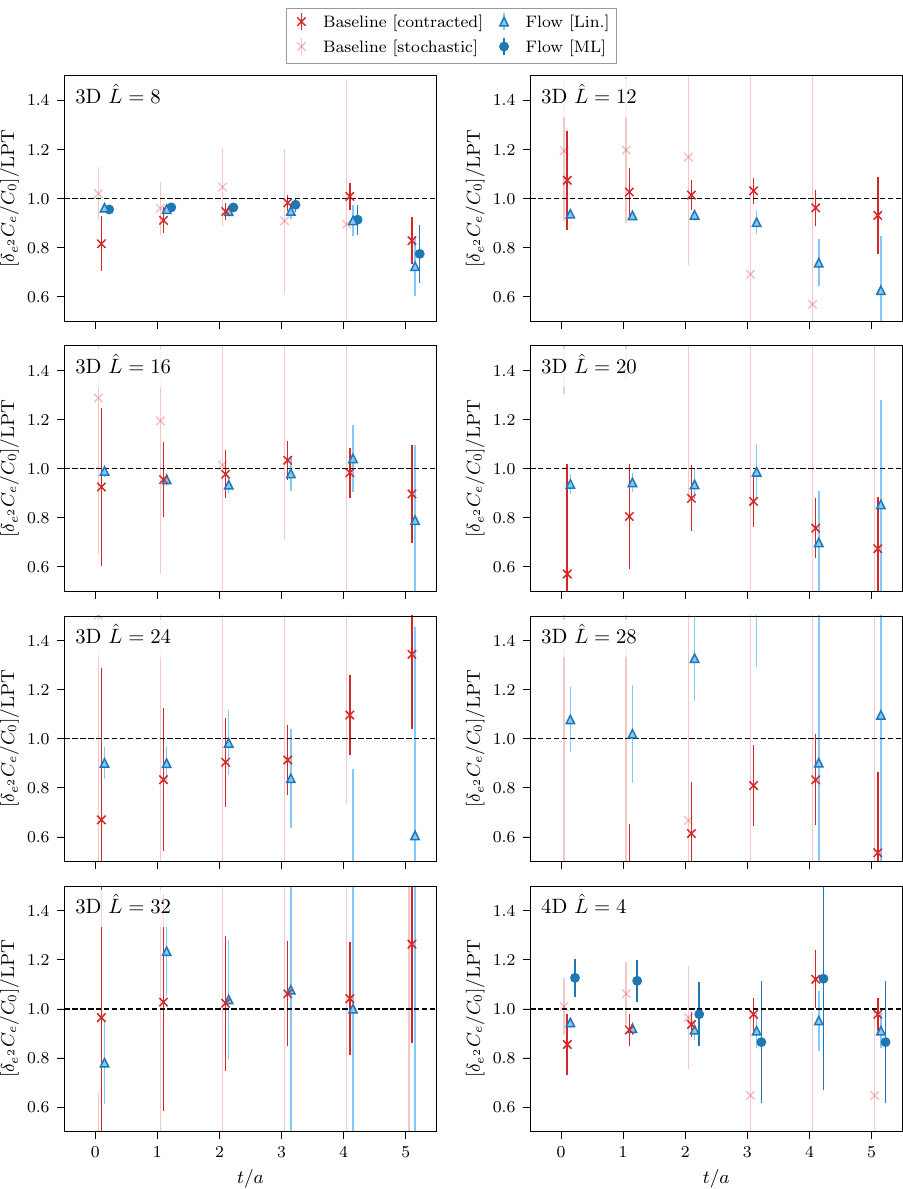}
    \caption{The $O(e^2)$ effective mass shift divided by the analytical LPT result in $N_d = 3, 4$ for various $\hat{L}$. Details are identical to Fig.~\ref{fig:effectivemass_varyL_2D_e2}. \label{fig:effectivemass_varyL_3D_e2}}
\end{figure}

We compare this $O(e^2)$ evaluation to the fixed-order RM123 approach~\cite{deDivitiis:2013xla}.
For an expectation value starting at order $e^2$, one has to leading order
\begin{align}\label{eq:rm123exp}
\obspe = \obsiso + \obsprime 
\, ,
  \quad \text{where} \quad
    \sprime = - e^2 \, \sintdern{2} + \frac{e^2}{2}\, \sintdern{1}^2 \, ,
\end{align}
which extracts the $O(e^2)$ contributions generated by the interaction in the Euclidean path integral.
Here the subscript $c$ refers to the vacuum-subtracted contribution. 
Note that if $\mathcal{O}$ itself depends on $e$, unlike for the two-point function studied here, there will be an additional term in the expansion. 
For the fixed-order RM123 approach, we use two different methods for the order-$e^2$ matrix element on the right-hand side of Eq.~(\ref{eq:rm123exp}). In the first, the matrix element is sampled over $S_0(\phi)+S_A(A)$. In the second, a Wick contraction over the two photon fields is performed, leaving an observable involving the exact photon propagator. The former is a standard approach to perform fixed-order RM123 calculations~\cite{deDivitiis:2013xla,Boyle:2017gzv,Davoudi:2018qpl}, while the latter is tractable via an exact Fourier transform in scalar QED. In Figs.~\ref{fig:effectivemass_varyL_2D_e2}--\ref{fig:effectivemass_varyL_3D_e2} we show results comparing these methods to the machine-learned and linear flows. All results are plotted as a ratio versus the exact LPT result derived in \eqref{eq:lpt-corr}.
The results for the flow method are obtained by the finite-difference estimate
\begin{equation}
    \delta_{e^2} C_e / C_0 \approx \frac{1}{2 e^2} \frac{C_{e} + C_{-e} - 2 C_0}{C_0},
\end{equation}
which introduces a bias arising from corrections at $O(e^4)$ and higher.
While finite difference uncertainties can be arbitrarily reduced by training flows for smaller values of the coupling, here our primary objective was the extraction of all-orders effects, and as a result finite-difference uncertainties in these estimates are notable. In future work, it would also be interesting to perform an exact expansion of the trained flows, as in Ref.~\cite{Abbott:2026ylv}, which could yield estimates without these effects.

\end{document}